\pdfoutput=1
\documentclass[11pt]{article}

\PassOptionsToPackage{numbers, compress}{natbib}

\usepackage[final]{neurips_2025}

\title{Synthesizing Performance Constraints for Evaluating and Improving Code Efficiency}

\author{
Jun Yang, Cheng-Chi Wang, Bogdan Alexandru Stoica, Kexin Pei \\
  \texttt{\{juny,casperwang,bastoica,kpei\}@uchicago.edu} \\
  Department of Computer Science, The University of Chicago
}
\usepackage{microtype}
\usepackage{graphicx}
\usepackage{comment}
\usepackage{subfig}
\usepackage{booktabs} % for professional tables
\usepackage{multirow}
\usepackage{colortbl}
\usepackage{tabularx}
\usepackage{ulem}
\usepackage[dvipsnames]{xcolor}

\usepackage{times}
\usepackage{latexsym}
\usepackage[T1]{fontenc}
\usepackage[utf8]{inputenc}
\usepackage{inconsolata}

\usepackage[hidelinks,colorlinks=true, linkcolor=blue, citecolor=blue, urlcolor=blue]{hyperref}

\usepackage{natbib}
\renewcommand{\cite}{\citep}

\usepackage{amsmath}
\usepackage{amssymb}
\usepackage{mathtools}
\usepackage{amsthm}
\usepackage{xspace}
\usepackage{pifont}
\usepackage{caption}
\usepackage{tikz}
\usepackage[textsize=tiny]{todonotes}
\usepackage[dvipsnames]{xcolor}

\usepackage[capitalize,noabbrev]{cleveref}
\crefformat{section}{\S#2#1#3}
\crefformat{subsection}{\S#2#1#3}
\crefformat{subsubsection}{\S#2#1#3}

\usepackage{wrapfig}
\usepackage{listings}
\usepackage{inconsolata}

\definecolor{codebg}{rgb}{0.97,0.97,0.97}
\definecolor{commentgray}{rgb}{0.3,0.5,0.3}
\definecolor{keywordblue}{rgb}{0.2,0.2,0.7}
\definecolor{stringorange}{rgb}{0.6,0.3,0.1}

\definecolor{numbergreen}{rgb}{0.0,0.5,0.0}

\lstdefinestyle{fancy}{
    backgroundcolor=\color{codebg},
    commentstyle=\color{commentgray}\ttfamily\itshape,
    keywordstyle=\color{keywordblue}\bfseries,
    stringstyle=\color{stringorange},
    numberstyle=\tiny\color{gray},
    basicstyle=\scriptsize\ttfamily,
    breaklines=true,
    numbers=left,
    numbersep=3pt,
    showstringspaces=false,
    tabsize=2,
    frame=single,
    rulecolor=\color{gray!40},
    xleftmargin=0.5em,
    framexleftmargin=0.5em,
    captionpos=b,
    literate=
        *{0}{{{\color{numbergreen}0}}}{1}
         {1}{{{\color{numbergreen}1}}}{1}
         {2}{{{\color{numbergreen}2}}}{1}
         {3}{{{\color{numbergreen}3}}}{1}
         {4}{{{\color{numbergreen}4}}}{1}
         {5}{{{\color{numbergreen}5}}}{1}
         {6}{{{\color{numbergreen}6}}}{1}
         {7}{{{\color{numbergreen}7}}}{1}
         {8}{{{\color{numbergreen}8}}}{1}
         {9}{{{\color{numbergreen}9}}}{1}
}
\usepackage{makecell}

\usepackage{pifont}

\usepackage{placeins}

\theoremstyle{plain}

\theoremstyle{definition}

\theoremstyle{remark}

\definecolor{CustomPurple}{HTML}{58508d}
\definecolor{CustomViolet}{HTML}{bc5090}
\definecolor{CustomOrange}{HTML}{ff6361}
\definecolor{CustomYellow}{HTML}{ffa600}

\newcommand{\bench}{\textsc{PerfForge}\xspace}

\newcommand{\tool}{\textsc{Wedge}\xspace}

\newcommand{\para}[1]{\smallskip\noindent{\textbf{#1}}}

\newcommand{\ie}{i.e., }
\newcommand{\eg}{e.g., }
\newcommand{\etc}{etc.,\xspace}

\newcommand{\hytt}[1]{\texttt{\hyphenchar\font=\defaulthyphenchar #1}}
\newcommand{\codeIn}[1]{\hytt{#1}}
\newcommand{\llmfull}{Large Language Model\xspace}
\newcommand{\llm}{LLM\xspace}

\newcommand{\llmsymbol}{\codeIn{LLM}\xspace}

\newcommand{\codecontest}{CodeContests\xspace}
\newcommand{\codeforces}{Codeforces\xspace}

\newcommand{\pcc}{performance-characterizing constraints\xspace}
\newcommand{\pccsmall}{performance-characterizing constraints\xspace}

\newcommand{\lengthstressing}{length-stressing\xspace}

\newcommand{\website}{\url{https://github.com/UChiSeclab/perfforge}}

\newcommand{\tgprompt}{TG-prompt\xspace}
\newcommand{\evalperf}{\textsc{EvalPerf}\xspace}
\newcommand{\evalperfslow}{\textsc{EvalPerf\textsubscript{slow}}\xspace}
\newcommand{\evalperfrandom}{\textsc{EvalPerf\textsubscript{rand}}\xspace}
\newcommand{\perffuzz}{PerfFuzz\xspace}

\newcommand{\effilearner}{\textsc{Effi-Learner}\xspace}

\newcommand{\pie}{\textsc{Pie}\xspace}
\newcommand{\piehq}{\textsc{PIE\textsubscript{h}}\xspace}
\newcommand{\piecond}{\textsc{PIE\textsubscript{c}}\xspace}
\newcommand{\pieuncond}{\textsc{PIE\textsubscript{a}}\xspace}
\newcommand{\afl}{AFL++\xspace}

\newcommand{\wedgenoinstrument}{\textsc{Wedge\textsubscript{NoInstr}}\xspace}
\newcommand{\wedgedefaultmutator}{\textsc{Wedge\textsubscript{DefaultMut}}\xspace}
\newcommand{\wedgeafl}{\textsc{Afl++}\xspace}

\newcommand{\defaultslowfive}{CC\textsubscript{slow}\xspace}
\newcommand{\defaultall}{CC\textsubscript{default}\xspace}
\newcommand{\wedgeslowfive}{PerfForge\xspace}

\newcommand{\noprofile}{None\xspace}
\newcommand{\publicfiveprofile}{CC\textsubscript{default}\xspace}
\newcommand{\wedgeprofile}{PerfForge\xspace}

\usepackage{xcolor}
\newcommand{\placeholder}[1]{\colorbox{green!20}{\texttt{#1}}}

\newcommand{\numtodo}[1]{#1\xspace}

\begin{document}

\maketitle

\begin{abstract}
Large Language Models (LLMs) have been increasingly used to optimize code efficiency.
Evaluating their effectiveness and further suggesting optimization opportunities often rely on high-quality tests to demonstrate the performance bottlenecks presented in the program.
However, existing approaches rely on a limited set of hand-curated inputs or LLM-generated uninteresting \lengthstressing tests, failing to reveal more nuanced optimization opportunities.
We present \tool, a framework for generating performance-stressing input given the program under test.
\tool synthesizes explicit performance-characterizing constraints in the form of branch conditions to partition the programs' execution space into performance-specific regions.
When integrated with the coverage-guided fuzzer, reaching different regions introduces explicit rewards for test generation to explore inefficient implementations.
% Our framework relies on the fuzzy code-comprehension capabilities of LLMs to generate performance constraints as an intermediate reasoning step.
% These constrains are then integrated into the target programs as branch instructions that a fuzzing tool can explore.
Our evaluation shows that \tool introduces a significant slowdown compared to the tests in CodeContests and those claimed to be optimized by existing approaches.
From the utility perspective, integrating our tests substantially improves the existing code optimization approaches that rely on test-driven execution feedback.
We release \bench, the performance tests generated by \tool, to benchmark future approaches for efficient code generation at \website.
\end{abstract}

\iffalse
\begin{abstract}
Large Language Models (LLMs) have been increasingly used in code generation, code optimization, and more. Existing benchmarks either focus on evaluating code correctness with a limited set of simplistic test inputs, which cannot effectively reveal code efficiency issues, or rely on synthesizing input generators via prompting to generate large \lengthstressing inputs, failing to reveal more nuanced optimization opportunities.
\end{abstract}
\fi

\section{Introduction}

Large Language Models (LLMs) have shown intriguing promise in optimizing code efficiency beyond compiler techniques~\cite{berger2020scalene, cummins-arxiv24, huang-neurips24, liu2024learning, shypula-iclr24, li2025editlord, peng2024perfcodegen, du2024mercury, huang2024effi}.
Evaluating the effectiveness of these LM-based code optimizations relies on performance-stressing tests. 
For example, an optimization from recursion to iteration in Fibonacci number calculation incurs only a negligible performance improvement when evaluated with a default test ($n=3$) that focuses on testing correctness, while a performance-stressing input ($n=40$) reveals the orders ($10^{6}$) of the larger gap.
Moreover, as some approaches integrate execution feedback to further optimize the code~\cite{huang-neurips24, peng2024perfcodegen, zelikman2024self}, running performance-stressing tests reveals more precise optimization opportunities by exposing performance bottlenecks.

% Precisely evaluating code efficiency is critical, since imprecise evaluation may give misleading conclusions on whether and how much the \llm can optimize code, and feedback-based code optimization techniques may also suffer from imprecise execution feedback. 
% Therefore, performance-stressing tests are critical to evaluate the efficacy of the optimization and further suggest optimization opportunities.
% 
% 

% Imprecise evaluation may give misleading conclusions on whether and how much the \llm can optimize code.
% 
% In addition, code optimization techniques may require execution profiling feedback, \eg execution time or memory usage to iteratively refine the code~\cite{huang-neurips24}. High-quality, performance-stressing tests are playing an important role here, because simplistic test executions are more susceptible to noise from testbeds and incur high variation in repetitive executions.  shows that the smaller executions are more sensitive to system noise despite using a clean test bed.

% \para{Limitations of existing approaches.} 
% 
% Unfortunately, most existing works \cite{huang-neurips24-effibench,shypula-iclr24} evaluating code efficiency still rely on simplistic  provided by the platform or dataset that aim to test correctness, not efficiency. 
Unfortunately, most existing code optimization approaches still leverage correctness tests to evaluate and suggest optimizations~\cite{huang-neurips24, shypula-iclr24, peng2024perfcodegen}.
% However, the correctness and performance represent the two orthogonal code properties.
However, the correctness tests alone are often insufficient to expose the inefficient code implementation.
For example, existing tests in the common benchmarks, \eg HumanEval~\cite{chen2021evaluating} has been shown to have limited scope and low complexity and thus fail to adequately stress the code performance against more demanding conditions~\cite{liu-colm24}. 
% since they were created to test correctness instead of performance .
As a result, they are also more susceptible to the noise introduced in the execution environment, thus failing to reliably quantify the optimization and reveal insightful optimization opportunities.

To generate performance-stressing tests, recent works have started to leverage LLMs by prompting them to generate test generators~\cite{liu-colm24}.
% However, the \llm-based input generator tends to resort to simply generating large, \lengthstressing inputs without reasoning about more nuanced requirements (despite CoT few-shot learning being used), failing to stress programs where more fine-grained inputs are necessary, as shown in Section~\ref{sec:quantitative_results}.
For example, EvalPerf~\cite{liu-colm24} introduced a scale parameter to control the input size, with the assumption that it is the key determining factor for performance-stressing.
However, such a biased preference over large tests misses the opportunity to reason about their relationship to inefficient program behaviors beyond the size.
% For example, calling quicksort can suffer from the suboptimal performance~\cite{pestios_ccs2017} when its input is reversely sorted ($O(n^2)$ in the worst case), while the input length is less interesting as it is merely bounded by the algorithmic complexity or the task specifications, \eg the constants defined in the problem description~\cite{liu-neurips23, li-science22}.
For example, calling quicksort can suffer from the suboptimal performance~\cite{pestios_ccs2017} when its input is reversely sorted ($O(n^2)$ in the worst case).
When two inputs are both at the maximum length \codeIn{n}, the reversely sorted one is more stressing than another randomly ordered one ($\mathcal{O}(n\log n)$) on average.

% , failing to stress programs where more fine-grained inputs are necessary.
% The nuanced requirements may involve diverse patterns, \eg relationship among input elements, structure of inputs, \etc, that solely increasing the size of the generator inherently cannot cover.

% or fuzzing with the execution time as the feedback besides the code coverage~\cite{lemieux_issta2018, pestios_ccs2017}
%Alternatively, the approaches based on fuzzing rely on the execution time as the feedback at each fuzzing round to guide the input search.
% This incurs substantial performance overhead and lacks a more fine-grained oracle for measuring the nuanced inefficient implementation.

% Coverage-guided search-based test input generation~\cite{fraser2011evosuite,fioraldi-woot20} is a better fit for this role.
% Coverage-guided test generation partitions the input space by path coverage, and continuously mutate the existing inputs to search uncovered input space.
% By exhaustively searching different partitions of the input space, different input patterns can be covered, including the diverse stressing input patterns.

% 
% .
% 

% In addition, PerfFuzz relies on AFL default mutators (bitflip, byteflip, crossover \etc), thus struggling to generate inputs that conform to the validity constraints, which further reduced the efficiency of fuzzing.

% \para{Our approach.} To bridge this gap, we propose \tool, a framework that combines performance constraints reasoning and coverage-guided fuzzing to adaptively generate inputs that stress the programs the most. 

\para{Our approach.}
We present \tool, a framework that generates performance test inputs beyond simply stressing the sizes.
Our key insight is that the limitation of LLMs in generating performance-stressing tests boils down to the inherent challenge of connecting the \textit{local} performance-related program behavior all the way back to the program inputs~\cite{jiang-ase24,xie2025corebenchmarkingllmscode}, while directly reasoning about the local behaviors is comparatively easier.
For example, we can easily specify the local variable \codeIn{arr}, the argument to a \codeIn{quicksort} deeply nested inside the program, to be \textit{reversely sorted} to trigger its inefficient behavior, while predicting what program inputs lead \codeIn{arr} to be reversely sorted is more challenging as that requires reasoning about the control and data flow based on the precise understanding of the program semantics.
Such reasoning is extremely challenging due to the overwhelming search space, \eg tracking a combinatorial number of program paths~\cite{cadar2008klee, cousot1996abstract, jiang-ase24, khedker2017data}.
% while we are not clear what kind of inputs leads to \codeIn{arr} being reversely sorted.

Based on such insight, \tool alleviates LLM reasoning on performance-related behavior by asking it to synthesize \textit{the performance-characterizing constraints} as condition checkers, \eg \codeIn{all(l[i] > l[i+1] for i in range(len(l)-1))}, and instrument the program with these checkers at the appropriate program points.
% We then employ coverage-guided fuzzers to explore the input space efficiently and systematically to find performance-stressing inputs.
% 
\tool then leverages the coverage-guided fuzzers, the search-based testing technique~\cite{fioraldi-woot20,fraser2011evosuite} with the goal to maximize the code coverage, to scale test input generation that sidesteps the expensive iterative queries to LLMs.
% Coverage-guided fuzzing is a mature and powerful search-based testing technique that has been shown to be effective in test generation and bug detection~\cite{fioraldi-woot20,fraser2011evosuite}, which is a better fit for searching constraint-satisfying inputs.
% It systematically searches the input space to generate inputs to maximize code coverage.
As the inputs achieving new coverage are rewarded and prioritized in the fuzzer, checker branches inserted by \tool serve as the coverage signal to bias the fuzzing to generate likely-stressing inputs more efficiently.

% To further support local constraint reasoning, employ (1) contrasting input, (2) (3) performance reasoning template.
% To further accelerate fuzzing, we employ reasoning of the enforce during and after fuzzing campaign, input grammar as constraint-conditioned input mutator, iterative self-refinement validator/contract generation (PBE)

% To further support performance constraint reasoning, 
% 
% \tool (1) employs a variant of contrastive COT prompting~\cite{chia_arxiv23} to obtain insight about what kind of inputs can make the program run slow , (2) designs a performance reasoning template that explicitly instructs the \llm to systematically reason about performance constraints in multiple phases and implement the corresponding constraint checkers.
% To help fuzzing to find performance-stressing inputs efficiently, \tool (1) instruments the program by inserting the constraint checker branches to bias the reward of fuzzing (\eg covering uncovered branches) towards satisfying the inserted constraints, (2) generates a constraint-aware mutator by prompting the \llm with the performance constraints to steer the input mutation towards likely constraint-satisfying inputs.

\begin{figure*}[t!]
\includegraphics[width=\linewidth]{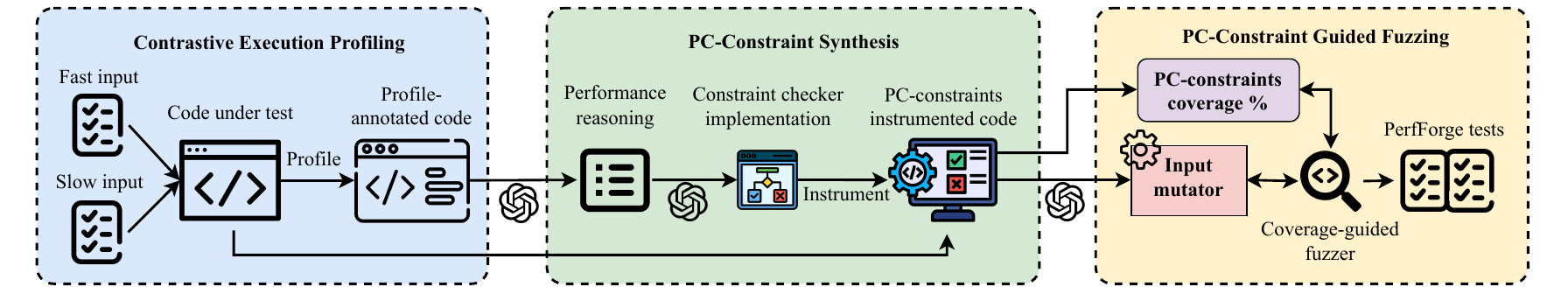}
\caption{Workflow of \tool. 
First, our tool profiles the code-under-test to identify a pair of inputs with contrastive execution profile (``fast'' vs ``slow'' execution). Second, with this information, it asks a \llm to infer performance-characterizing constraints and instrument the code with checkers. Third, it runs the instrumented code through a customized fuzzing tool to find performance-stressing inputs.}
%\vspace{-0.5cm}
\label{fig:wedge-workflow}
\end{figure*}

To enhance performance constraint reasoning, we develop a reasoning template that elaborates on the procedures to contrast the pair of disparate execution profiles to gain insight into inefficient implementations.
% what kind of inputs can make the program run slow by prompting a <\textit{slow}, \textit{fast}> input pair and their execution feedback.
% It then leverages high-quality in-context examples to generate performance-characterizing constraints.
We then instruct the \llm to reason about performance constraints (in natural language and code) in multiple phases to localize the appropriate program points and implement the corresponding constraint checkers.
% Based on the generated constraints and the corresponding program points, \tool instruments the program with the checker branches to bias the reward of fuzzing (\ie reaching uncovered branches) towards satisfying the inserted constraints.
% instead of relying on the AFL++ default mutators which perform arbitrary mutations such as bit flip, byte flip, \etc 
Besides guiding the fuzzer using constraint checkers, \tool further accelerates the input search by replacing the fuzzer's default input mutator~\cite{fioraldi-woot20} with a constraint-aware one that steers the input mutation towards likely constraint-satisfying inputs, while also enforcing the mutation to respect the input grammars~\cite{srivastava2021gramatron, liu2024oss-fuzz-gen, shi2024harnessing, zhang2024effective}.
% Such a mutator steers the input mutation towards likely constraint-satisfying inputs, further narrowing down the search space of performance-stressing inputs.
Figure~\ref{fig:wedge-workflow} presents our workflow (see Section~\ref{sec:framework} for details).

\paragraph{Results.}
Our extensive evaluation shows that the tests generated by \tool are substantially more performance-stressing than the default ones in the existing benchmark and those generated by the state-of-the-art techniques~\cite{liu-colm24,lemieux_issta2018} by 84.5\% (vs. \evalperfslow).
With more stressing tests, \tool precisely pinpoints the potential inefficient implementations and thus introduces approximately 10 percentage points more efficiency improvement on the generated code than that of default tests when used to guide the iterative code optimization approaches via test-driven execution feedback~\cite{huang-neurips24}.
% 9.42% = 49.31% - 39.89%
% , demonstrating the superiority of performance stressing tests in evaluating and improving LLM-based code optimization techniques.
% 
Our ablations confirm the effectiveness of the synthesized constraints in guiding the fuzzing and input mutation, \ie achieving 4$\times$ improvement over plain fuzzing using \afl.
In addition, we show that the generated constraints effectively characterize the performance, where the constraint-satisfying inputs are 38.6$\times$ slower than constraint-violating inputs.

\section{Overview}
\label{sec:overview}
% previous overview

We start with an overview of existing works on code efficiency evaluation and stress test generation.
We then use an example to demonstrate the advantage of \tool over the existing approaches.

\subsection{Benchmarking Code Efficiency and Performance-Stressing Test Generation}
\label{sec:overview-relatedwork}

% \para{Benchmarking Code Efficiency.}
\begin{comment}
Most existing code generation benchmarks focus on the correctness aspect.
For example, APPS~\cite{hendrycks2021measuring}, MBPP~\cite{austin2021program}, HumanEval~\cite{chen2021evaluating} and EvalPlus~\cite{liu-neurips23} work on curating high-quality Python benchmarks including coding problems along with correctness tests.
HumanEval-X~\cite{zheng2023codegeex}, MultiPLe~\cite{cassano2022multipl}, and MBXP~\cite{athiwaratkun2022multi} extend the Python tasks into other programming languages.
\end{comment}

While traditional code generation primarily 
    focused on generating correct 
    code~\cite{jain2024livecodebench, hendrycks2021measuring,austin2021program,chen2021evaluating,liu-neurips23,zheng2023codegeex,athiwaratkun2022multi,li-science22}, there are growing 
    efforts to generate efficient code beyond 
    correctness~\cite{shypula-iclr24,liu-colm24,huang-neurips24-effibench,peng2025coffecodeefficiencybenchmark}.
However, existing efficient code generation 
    techniques still largely rely on correctness 
    tests to evaluate the performance 
    improvement~\cite{shypula-iclr24, peng2024perfcodegen, waghjale-etal-2024-ecco}, which cannot faithfully measure the 
    performance improvement~\cite{liu-colm24, peng2025coffecodeefficiencybenchmark, huang-neurips24-effibench, huang2024effi}.
Some of them rely on the execution feedback to further optimize the code~\cite{peng2024perfcodegen, huang2024effi}.
These approaches can miss optimization opportunities when the tests do not reveal the performance bottleneck (see Section~\ref{sec:tests-utility}).

To address these challenges, recent works have focused on performance test generation to benchmark efficient code generation~\cite{lemieux-icse23, kang-icse23, deng2024large, huang-neurips24-effibench,peng2025coffecodeefficiencybenchmark,du2024mercury,liu-colm24}.
However, these approaches either generate infeasible inputs that do not stress and thus rely on manual correction, or their task formulation often prevents the LLM from reliably reasoning about the program behavior, \ie by directly prompting the LLM to generate the stressing inputs for the entire long-spanning program.
% , which often involves tracking multiple complex control and data flows.
% However, these approaches either generate 
    % infeasible or inefficient inputs and thus rely on manual correction, or their task formulation often prevents the LLMs from reliably reasoning about the program behavior, i.e., by directly prompting them to generate the stressing inputs purely based on the code snippets.
    % \edit{This often involves tracking and reasoning over multiple nontrivial control and data flows in the holistic long-spanning program, which LLMs have been shown to be not good at~\cite{liu2025codemindevaluatinglargelanguage,jiang-ase24}.}
    % ~\edit{NOTE: Kexin, Jun: I'm not sure what this means here. Do you want to say that these tools provide too much code to LLMs? I think this needs a bit of clarifying}.
    % 
With such a nontrivial task, LLMs have to identify the inefficient implementation, understand the run-time behavior to exercise it, and reason all the way to program inputs. 
Therefore, they often end up taking 
    ``shortcuts'' and reduce to only generating 
    \lengthstressing inputs that fail to reveal 
    more intricate inefficient 
    implementation.

% Besides benchmarking the efficiency of coding problem solutions, GSO~\cite{shetty2025gsochallengingsoftwareoptimization} generates performance tests for real-world repositories and workloads by prompting an LLM with the performance-optimizing code commit.
% It shares the high-level idea of direct prompting but often requires more challenging inter-procedural reasoning for test generation across longer context~\cite{liu-colm24,peng2025coffecodeefficiencybenchmark,huang2024effi,du2024mercury}.
% % \tool is complementary to GSO~\cite{shetty2025gsochallengingsoftwareoptimization} by generating performance-characterizing tests for different commits.
%  by decomposing the test generation task--targeting localized performance behaviors through explicit performance-characterizing constraint reasoning and exploitation.
% If extended to repositories-level, \tool can be expected to produce performance tests that expose inefficiencies at the level of individual commits.}
In addition to performance benchmarking using competitive-programming level code,
% code where no heavy-weight, long-context, inter-procedural reasoning is required, 
GSO~\cite{shetty2025gsochallengingsoftwareoptimization} extends the evaluation to repository-level and real-world workloads by prompting an LLM with the performance-optimizing commit.
It shares the high-level idea of direct prompting but requires more challenging inter-procedural analysis across a much longer context~\cite{jimenez2023swe, ding2023crosscodeeval}.
\tool complements the direct prompting approaches for performance testing by decomposing the test generation into local code behavior reasoning and efficient input search.
% While \tool is evaluated on single programs, it can be extended to repository-level, reasoning inefficiencies at the level of individual commits and searching for inputs that expose the inefficiencies.

% \para{Performance-Stressing Input Generation.}
% 1) the included additional tests are not intended to be stressing~\cite{shypula-iclr24,li-science22}.
% PIE~\cite{shypula-iclr24} includes test cases from AlphaCode~\cite{li-science22} generated with a fine-tuned \llm in addition to default tests in CodeNet~\cite{puri2021codenet}, but the AlphaCode tests are not intended to be stressing.
% First, the \llm-generated test cases often suffer from , introducing human effort and potential evaluation bias.
% \citet{huang-neurips24-effibench} prompts to generate a test case generator that can produce diversified test cases, but the generated test cases often suffer from low accuracy and require manual correction, introducing human effort and potential evaluation bias.
% Second, when generating stressing tests, the \llm often fails to reason how to make the program slow and take the shortcut, \ie only generating \lengthstressing inputs.
% \citet{liu-colm24} proposes to prompt an \llm with CoT few-shot learning to produce a test input sampler controlled by a parameter \textit{scale}.
% They use exponential input sampling to generate challenging but computable inputs.
% \evalperf mainly generates \lengthstressing inputs, but falls short when more fine-grained requirements are necessary, as shown in our results Section~\ref{sec:quantitative_results} and prior studies~\cite{zhang_ase2011}.

Performance testing has been studied extensively before \llm-based approaches based on symbolic execution and fuzzing~\cite{pestios_ccs2017,lemieux_issta2018,zhang_ase2011,burnim_icse09,chen_icse16}.
% ~\bo{I think we are too broad and too vague here. by mentioning `static' and `dynamic analysis'. Instead we should be explicit. 
% We should point out that the tools we cite either use symboic execution or fuzzing. In fact, I'd say that these two are the most common ``input generation'' techniques. I feel everything we cite here is actually fuzzing.}
% \edit{However, many of these techniques 
%     need to exhaustively explore a large program path space,
%     intensively collect a significant number of execution 
%     traces, heavily rely on manually-crafted heuristics
%     
%     % , 
%     % incur high profiling overhead
%     , or lack 
%     test oracles to precisely capture 
%     inefficient behaviors or symptoms.}
While the former can suffer from poor scalability, the latter can also incur expensive instrumentation for collecting and parsing profiling information, \eg customized coverage metrics and scheduling algorithms to guide fuzzing and input search, and can lack test oracles to precisely capture inefficient behaviors or symptoms.

% We resolve the above challenges by identifying and exploiting the performance-characterizing constraints in the program.
\tool restricts LLMs to focus exclusively on local \pcc to avoid relying on the LLMs to reason globally about the input, while leveraging efficient input search from fuzzing.
Therefore, it captures the inefficient behaviors without expensive profiling and explicitly encourages test generation towards reaching inefficient implementation beyond being \lengthstressing.

\subsection{Motivating Example}
\label{sec:overview-example}

\begin{figure*}[!t]
\includegraphics[width=\linewidth]{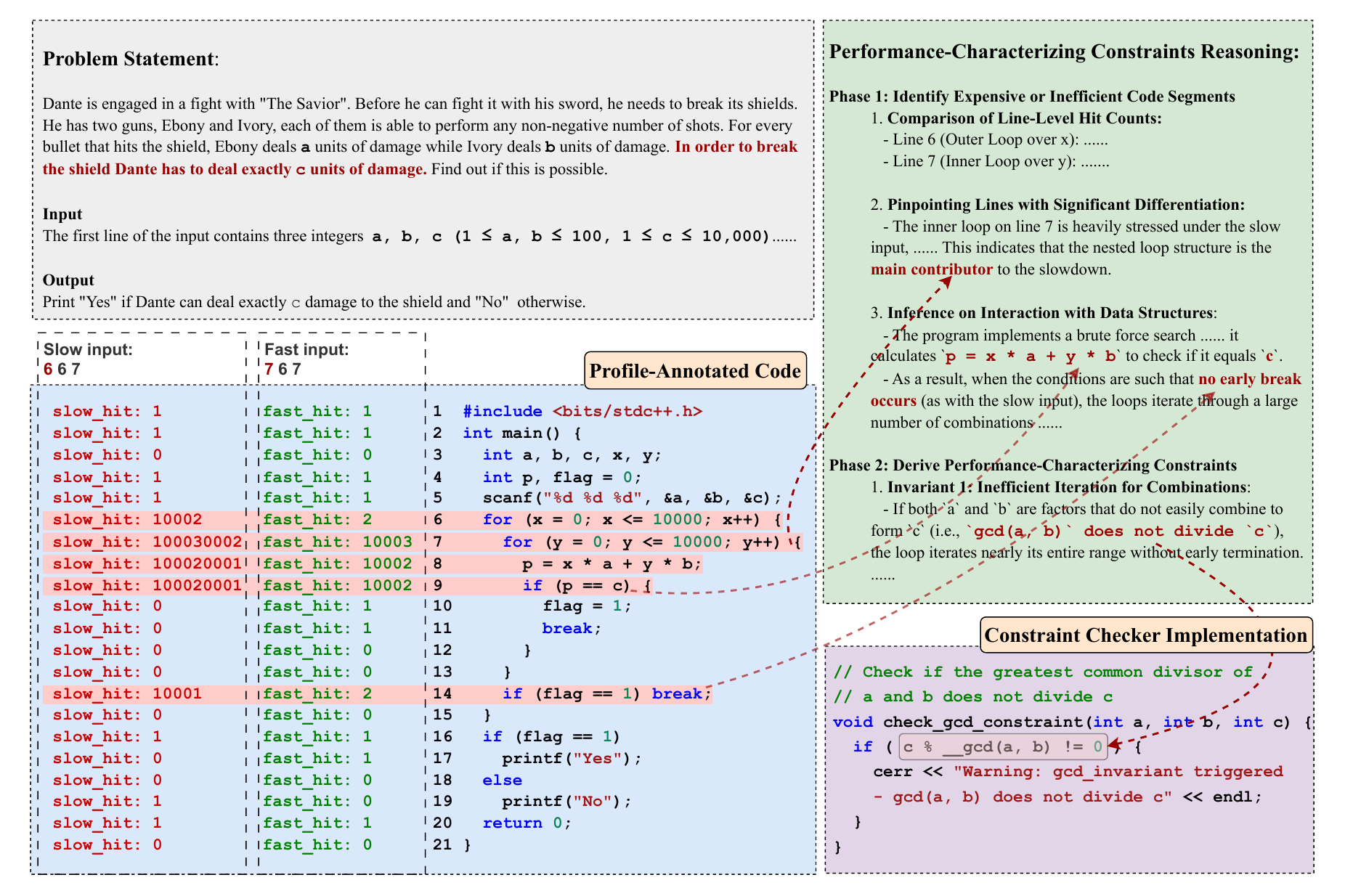}
\caption{Motivating example from \codeforces (prob. 633A, sol. 622) showing how \tool reasons about and generate \pccsmall, and implements corresponding checkers.}
\label{fig:motivating}
\end{figure*}

Let us consider an example from CodeContests, \codeforces problem 633A~\cite{codecontests-p633A} (gray box,~\cref{fig:motivating}).
Given three integers $a$, $b$, and $c$ ($1\leq a, b\leq100, 1\leq c \leq 10,000$), the goal is to decide whether there exist non‑negative integers $x$, $y$ such that $a\cdot x+b\cdot y=c$.
This problem is classically known as finding solutions to a two‑variable linear Diophantine equation~\cite{wiki:Diophantine_equation}.

The code snippet (blue region) shows an implementation that solves this problem.
The code systematically tries every pair of values by iterating two nested loops over fixed upper bounds of $10,000$ (lines 6–7), computing the value of the Diophantine equation, and checking whether it is equal or exceeds $c$.
Because the loops use fixed upper bounds rather than adapting to the value of $c$, the code could examine nearly the entire $10,000^2$ value space.
Apart from skipping sums greater than $c$ or breaking once a match is found (lines 9-11), the code bears the full brute‑force cost.

The right half of~\cref{fig:motivating} (green and purple boxes) shows how \tool infers \pcc specific to this program, inspired by the contrasting execution traces that share similar inputs but have disparate behavior (manifested by the per-statement execution counts).
In particular, our tool identifies specific relations among the local variables \texttt{a,b,c} to stress the nested loops to exhaust their maximum iterations.
The green box shows the \llm's reasoning process, while the purple box shows the \pcc synthesized as a C++ checker by the \llm to be instrumented in the program.

% We provide a more in-depth analysis of how \tool's inference in~\cref{appx:motivating-example}.
% Our key observation is that these constraints are more local, fine-grained, easier to generate, and cannot be captured by state-of-the-art techniques (e.g., \cite{liu-colm24,peng2025coffecodeefficiencybenchmark}, \cite{lemieux_issta2018}), which focus primarily on maximizing the input values and size.
% patch_1
Our key observation is that these constraints are more local, fine-grained, easier to generate, and cannot be captured by state-of-the-art techniques (e.g., \cite{liu-colm24,peng2025coffecodeefficiencybenchmark,lemieux_issta2018}), which focus primarily on maximizing the input values and size.
Therefore, such performance-stressing constraints serve as more appropriate interfaces for LLMs to communicate their reasoning to the existing test generation tools than directly asking them to generate performance-stressing inputs.
    
% Our intuition is that by reasoning local \pcc, we can obtain distilled knowledge about performance bottlenecks. 
% The distilled knowledge can be embedded into the original program in the form of constraint checker branches to guide the off-the-shelf fuzzing to exhaustively search for the constraint-satisfying inputs, i.e., likely performance-stressing inputs.
\section{\tool Framework}
\label{sec:framework}

% We formally define the performance-stressing test generation problem.
% We then describe the key components in \tool (shown in Figure~\ref{fig:wedge-workflow}).

% This section elaborates on the key components in \tool as presented in Figure~\ref{fig:wedge-workflow}.

\paragraph{Problem Statement}
We formally define the performance-stressing test generation problem.
Given a program $\mathcal{P}$ accepting a valid input (conforming to a validity specification $\mathcal{V}$)
% , details of checking $\mathcal{V}$ described in Appendix~\ref{sec:validator})
, the set of all valid inputs is denoted as $\mathcal{I_V}$.
With a valid input $\forall i \in \mathcal{I_V}$, the execution of $\mathcal{P}$ (denoted as $E_i=\mathcal{P} \cdot i$) yields an execution time~\footnote{For clarity, we use ``execution time'' here to represent the execution cost, but in experiments we use the number of CPU instructions. See our justification in Section~\ref{sec:setup}} $T_i$.
The goal of stress test generation is to generate a subset of valid inputs $I^* \subset \mathcal{I_V}$, such that the average execution time of $I^*$ is maximum.
% It is challenging to generate stress tests from mainly two aspects: 1) the difficulty of reasoning about what tests are stressing, 2) the huge search space.

At a high level, \tool takes as inputs a coding problem statement $\mathcal{S}$, a correct solution program $\mathcal{P}$, a set of default correctness tests $\mathcal{I}_D$, and an \llmfull (\codeIn{\llm}), and produces a set of performance-stressing test inputs $I$: $I=\tool(\mathcal{S},\mathcal{P}, \mathcal{I}_D, \llmsymbol)$.

\subsection{Contrastive Execution Profiling }
\label{sec:exec-feedback-collection}

We first collect high-quality contrastive execution feedback from fast and slow executions to facilitate reasoning about \pcc.
This is achieved in two steps.

\para{Contrastive input pair mining.} 
\tool runs $\mathcal{P}$ against a set of user-provided tests $\mathcal{I_D}$, \eg existing correctness tests provided by the dataset to mine a contrastive (slow, fast) input pair $(i_{slow}, i_{fast})$.
Note that $i_{slow}$ does not have to be performance-stressing; the pair needs only to provide contrastive diagnostic hints to help localize potentially inefficient implementation, i.e., the performance bottleneck.
% While running these tests, \tool collects line-level execution profiles (\eg, coverage and hit count) and the number of executed assembly instructions.
% We use number of instructions, denoted as $|I|$, to reflect the cost of executions, as instruction count is more reliable and less susceptible to noise as shown in Section~\ref{sec:quantitative_results}.
% While running these tests, \tool collects the number of executed hardware instructions.
% We use number of instructions, denoted as $|I|$, to reflect the cost of executions, as instruction count is more reliable and less susceptible to noise as demonstrated by \citet{liu-colm24}.
% While running these test inputs, \tool collects the execution cost (\#instructions, denoted as $|I|$, in our experiments) of each input.

During test execution, \tool collects the execution cost of each input, measured by the number of executed instructions (denoted $|I|$ in our experiments).
% \tool pairs up the inputs in twos, and calculate the similarity and execution cost ratio of the two inputs.
% The similarity is the sum of \textit{match ratio} (number of common array elements / length of the shorter input array) and \textit{Jaccard similarity}~\cite{niwattanakul2013using}.
% The execution cost ratio is the ratio of execution cost of slow input to that of fast input.
\tool then mines the contrastive input pairs based on the two metrics: (1) similarity defined as the sum of the match ratio (i.e., the number of common array elements divided by the length of the shorter array) and the Jaccard similarity~\cite{niwattanakul2013using}, and (2) execution cost ratio, defined as the ratio of the slow input’s cost $|I|_{slow}$ to that of the fast input $|I|_{fast}$.
Input pairs are then ranked based on their similarity and execution cost ratio, and \tool selects the top-ranked pair as the contrastive input pair $(i_{slow},i_{fast})$.
% For input pairs with the same similarity, we further rank them by execution cost ratio.
% 
% \sout{The insight of using contrasting input pairs to inspire constraint reasoning comes from \textit{delta debugging}~\cite{zeller-tse02}.
% Delta debugging isolates bugs by systematically comparing and narrowing down the difference (delta) between the passing case and the failing case, to find the minimized set of bug-inducing circumstances.
% By providing the similar-but-different input pairs along with the execution feedback, \llm can ``delta-debug'' by pinpointing the differences between inputs and feedback to understand why the slow input makes the program run slow.}

\para{Profiling feedback collection.}
% After identifying a potentially promising input pair, 
\tool executes $\mathcal{P}$ with ${i_{slow}}$ and ${i_{fast}}$, collecting execution feedback (coverage and hit count) $F_{slow}$ and $F_{fast}$.
Considering such a contrastive execution pair provides the key behavior insight~\cite{lin_icse17}, we prompt \llm to pinpoint the differences to reason why one input leads to significantly slower execution.
% The idea parallels classic debugging techniques like delta debugging~\cite{zeller-tse02} and statistical debugging~\cite{zheng_icml2006}, where developers analyze correlations across multiple input-result repairs to identify root causes.
% By presenting similar input pairs along with their respective execution feedback, \llm performs ``delta debugging'' by pinpointing the differences to reason why one input leads to significantly slower execution.
% Powered by \llm, our approach requires no human intervention thus can be fully automated.

\subsection{Performance-Characterizing Constraint Synthesis}
% In this stage, \tool generates the constraints (in natural language) in the first step.
% In the second step, it prompts the \llm to implement the constraint checkers based on the natural language constraints and insert them to the fuzz driver.
\tool generates the constraints in two steps: it initially generates the constraints $\mathbb{C}$ in natural language, then prompts the \llm to implement the corresponding constraint checkers and insert them to the fuzz driver $\mathcal{P}$ to produce the instrumented fuzz driver $\mathcal{P'}$.

\para{Performance-characterizing constraint reasoning.}
A constraint is a predicate on the program state (\eg variable values) and expressed as a conditional statement, \eg \codeIn{if (n > 1)}.
Given a performance-characterizing constraint $c$ and a given set of inputs $\mathcal{I_V}$, some inputs may satisfy the constraint while others may not.
We denote them as $\mathcal{I_S}$ and $\mathcal{I_N}$, respectively,
where $\mathcal{I_V}=\mathcal{I_S} \cup \mathcal{I_N}$, and the corresponding average execution time $\overline{\mathcal{T_S}}>\overline{\mathcal{T_N}}$.

\tool first constructs a comprehensive \textit{performance reasoning prompt template} that contains the problem statement $\mathcal{S}$, solution program $\mathcal{P}$, contrasting input pair $(i_{slow}, i_{fast})$, the profiling feedback information $F_{slow}$ and $F_{fast}$, and multiple manually-crafted constraints as few-shot examples.
The performance constraint reasoning technique can be denoted as: $\textit{ReasonPerf}(\llmsymbol, \mathcal{S}, \mathcal{P}, (i_{slow}, i_{fast}), (F_{slow}, F_{fast}))=\mathbb{C}$, where $\mathbb{C}=\left\{ c_i \right\}_{i=1}^N$ is a set of generated constraints.
The template explicitly instructs the \llm to reason about performance constraints in multiple phases, as shown in Figure~\ref{fig:motivating}.
In Phase 1, the \llm needs to identify expensive or inefficient code fragments.
This includes: 1) comparing line-level profiling information, \eg hit counts, between the fast and slow runs, 2) pinpointing lines or functions that get significantly more hits under the slow input, and 3) inferring how these lines might interact with data structures, loops, recursion, etc., especially as they relate to the input constraints (e.g., n <= 100).
In Phase 2, the \llm will derive performance-characterizing constraints in natural language.
By enforcing the \llm to reason about the constraints with Chain-of-Thought prompting~\cite{wei-nips22}, \tool collects insights into performance and generates high-quality constraints $\mathbb{C}$ (Figure~\ref{fig:motivating} green part).

\para{Constraint checker implementation.}
% \tool insert corresponding checker branches to the original program to produce the instrumented program.
% \tool writes down the constraints $c$ and the instrumented program $\mathcal{P'}$.
\tool prompts the \llm with the constraints $\mathbb{C}$ and instructs it to implement the checker code faithfully and produce the instrumented program.
The instrumented program with inserted checker code $\mathcal{P'}$, will be used as the target program to fuzz: $\mathcal{P'}=\textit{Instrument}(\llmsymbol, \mathcal{P}, \mathbb{C})$.

% \vspace{5em}
\subsection{Performance-Characterizing Constraint Guided Fuzzing}
\label{sec:constraing-guided-fuzzing}

In this stage, \tool launches coverage-guided fuzzing against the instrumented program $\mathcal{P'}$ to search for constraint-satisfying inputs.

\para{Constraint-aware mutator generation.}
\tool uses \afl as its fuzzing engine. 
% In this stage, given the instrumented program, \tool uses \afl as its fuzzing engine. 
% It, once again, relies on the \llm to generate mutators, given the instrumented program $\mathcal{P}$ and the natural language description of the performance-characterizing-constraints.
However, the default mutator of \afl (denoted as $\mathcal{M_D}$) targets at binary fuzzing (including operations like bitflip, byteflip, crossover, etc.), having no knowledge of input validity constraints, thus could generate mostly invalid inputs.
We implement a custom input-grammar- and constraints-aware mutator $\mathcal{M}_\mathbb{C}$ by prompting the \llm with mutator examples, problem statement $\mathcal{S}$ (\ie validity constraint $\mathcal{V}$), solution program $\mathcal{P}$, contrasting input pair $(i_{slow}, i_{fast})$, the profiling feedback information $(F_{slow}, F_{fast})$ and the generated performance constraints $\mathbb{C}$: $\mathcal{M}_\mathbb{C}=\textit{MutatorSyn}(\llmsymbol, \mathcal{S}, \mathcal{P}, (i_{slow}, i_{fast}), (F_{slow}, F_{fast}), \mathbb{C})$.

Mutator generation is more challenging than \evalperf~\cite{liu-colm24} and the input generator's generation~\cite{huang-neurips24,peng2025coffecodeefficiencybenchmark,du2024mercury}, as it has to be robust enough to make sure the mutated inputs follow the validity constraints and meanwhile as diversified as possible.
To resolve this challenge, \tool follows an iterative generate-and-fix fashion to ensure the robustness of mutators.
We put more details in Appendix~\ref{sec:mutator} due to the space constraints.
% 
% Note that this was already generated by the \llm in the previous stage.

% 
% For example, Listing~\ref{lst:p633a-constraint-checker} is another constraint checker for problem 633A in addition to 
% \codeIn{check\_gcd\_constraint} in Figure~\ref{fig:motivating}.
% Listing~\ref{lst:p633a-wedge} is a mutator generated for problem 633A.
% The mutator incorporates both validity constraints (\eg $1 \leq a, b \leq 100, 1 \leq c \leq 10000$) and performance constraints (\ie \codeIn{check\_close\_values\_constraint} in Listing~\ref{lst:p633a-constraint-checker}, \ie difference between a and b should be less than 5, either a or b should not be the divisor of c), thus can continuously produce diverse constraint-satisfying inputs in fuzzing.
% % While the generators generated by \evalperf or \directprompting fail to catch the complex performance constraints, as shown in Listing~\ref{lst:p633a-evalperfrandom}, Listing~\ref{lst:p633a-evalperfslow}, and Listing~\ref{lst:p633a-directprompting}.
% % 
% While the generator produced by the state-of-the-art technique \evalperf fails to catch the complex performance constraints, as shown in Listing~\ref{lst:p633a-evalperfrandom}.

\para{Constraint-guided fuzzing.}
Once mutators are generated, it launches a fuzzing campaign using the mutator $\mathcal{M}_\mathbb{C}$ on the instrumented program $\mathcal{P'}$, collecting all tests generated by fuzzer, \ie $\textit{CGF}(\mathcal{M_C},\mathcal{P'})=I$, where $I=\left\{ i_1,i_2,... \right\}$ are the fuzzer generated tests.
In the end, the tests generated by \tool form our benchmark \bench.
\section{Experiments}

% We evaluate \tool by (1) measuring the execution cost of our generated inputs (Section~\ref{sec:quantitative_results}), (2) investigating their utility at highlighting and addressing code inefficiencies (Section~\ref{sec:tests-utility}), and (3) conducting a comprehensive sensitivity analysis of \tool (Section~\ref{sec:sensitivity_analysis}).
% Due to space constraints, we include the description of our extensive evaluation details and exciting .

\subsection{Setup}
\label{sec:setup}

\para{Test generation baselines.}
We evaluate \bench tests (generated by \tool) against the following three representative baselines (two LLM-based and one fuzzing-based):
\textbf{\evalperf}~\cite{liu-colm24}, which uses LLMs to synthesize a parameterized input generator controlled by the input size parameter \textit{scale}.
Since \evalperf requires one canonical program as the reference implementation in the prompt, while our dataset has multiple ground-truth solutions per problem, we use the slowest and a randomly sampled solution as the reference implementation, forming two variants \evalperfslow and \evalperfrandom.
\textbf{\tgprompt}~\cite{du2024mercury,peng2025coffecodeefficiencybenchmark,huang-neurips24-effibench}, a direct prompting technique following recent works~\cite{du2024mercury,peng2025coffecodeefficiencybenchmark,huang-neurips24-effibench} which instructs an LLM to directly synthesize the performance test generator given the problem specification.
\textbf{\perffuzz}~\cite{lemieux_issta2018}, which is a state-of-the-art performance fuzzing tool that uses a performance-aware coverage metric that tracks the hit count of each control flow graph edge in the target program to search for inputs that either reach new edges or hit known edges more.

% We evaluate \bench tests against \edit{four baselines}, to compare the efficacy of our performance-stressing inputs. 
% % We use the first two to evaluate the efficacy of our performance-stressing inputs and the last two to determine their utility.
% %The first two serve to compare the efficacy of our performance-stressing inputs.
% We consider \evalperf~\cite{liu-colm24}, which uses LLMs to synthesize a parameterized input generator and progressively scales input size until a predefined timeout or out of memory.
% Since our dataset lacks canonical reference implementations, we consider two variants: \evalperfslow, which uses the slowest solution as the reference, and \evalperfrandom, which uses a random solution.
% \edit{We also compare with \perffuzz~\cite{lemieux_issta2018}, the state-of-the-art 
%     performance fuzzing tool that uses a 
%     multi-dimensional  to generate 
%     inputs that exercise cost-amplifying code paths.}
% We finally consider \tgprompt following the recent work ~\cite{du2024mercury,peng2025coffecodeefficiencybenchmark,huang-neurips24-effibench} that instructs an \llm to directly synthesize the performance test generator given the problem specification and its constraints. 
% 
% 
% 

\para{Utility baselines.}
To measure the utility of our generated tests, we consider two scenarios that \bench can help.
The first scenario is to provide execution feedback to help LLMs further optimize the code.
We consider \textit{\effilearner}~\cite{huang-neurips24}, an iterative code efficiency optimization based on test-driven execution feedback to guide the \llm in refining its generated code. 
% It iteratively generates a program implementing a particular specification, executes and captures overhead data which it next feeds back to the model.
The second scenario is to evaluate (ideally more precisely) existing code optimization approaches.
We consider running \bench against \pie~\cite{shypula-iclr24}, an \llm-based code optimization that finetunes the \llm on slow and fast code pairs, which relied on correctness tests to evaluate its performance improvements.

\para{Metrics.} 
We primarily rely on CPU instruction count to measure the effectiveness of \bench tests, considering it is more stable across runs, platform-agnostic, and strongly correlates with performance bottlenecks~\cite{curtsinger-sosp15,zhang-eurosys13,iorgulescu-atc18,delimitrou-asplos14}, while physical time is more prone to interference and noise~\cite{de2010new,zhang-eurosys13}.
% In contrast, instruction count is broadly acknowledged to be  more stable across runs, platform-agnostic and strongly correlate with performance bottlenecks~\cite{curtsinger-sosp15,zhang-eurosys13,iorgulescu-atc18,delimitrou-asplos14}.
It is also one of the key metrics for evaluating \llm-based code testing and optimization tools~\cite{liu-colm24, peng2025coffecodeefficiencybenchmark,shypula-iclr24} (more details in~\cref{appx:measurements}).
To further reduce the noise, we average the CPU instructions over \textit{five} runs for each program throughout all experiments.

\para{Dataset.} We evaluate \tool on CodeContests~\cite{li-science22} with a wide range of competitive programming problems and human‑written solutions.
Test cases include the default inputs from the original open-judge platforms as well as additional inputs generated by the authors~\cite{li-science22}.
We largely focus on C++ solutions to ensure comparable measurements, with a small subset of Python programs for the usefulness investigation (Section~\ref{sec:tests-utility}).
We rank the problems based on the coefficient of variation~\cite{liu-colm24} of the CPU instruction counts and select the top 300 problems.
This ensures the selected problems feature diverse solutions and potentially have enough room for optimizations for part of the solutions.
\tool generates tests for 207 of them, but after excluding those where baselines cannot produce valid inputs, we arrive at 154 problems and 33,020 C++ programs.

\paragraph{Fuzzing and input filtering.}
To collect inputs, We run \tool's fuzzing (based on our modified AFL++) for one hour for each solution in parallel.
Not all generated inputs strictly conform to the validity constraints $\mathcal{V}$ (Section~\ref{sec:framework}).
% \citet{liu-colm24} relies on manual contract annotation to enforce the validity constraint, which is known to be expensive and error-prone~\cite{brants-emnlp07}.
\tool applies a two-stage automatic filter to filter out likely invalid inputs.
\tool first prompts an LLM to generate the validator based on the problem statement and use the official tests in CodeContests to check the validity of the validator.
% The validator will check each generated input and discard invalid ones.
\tool then checks the output consistency across different solutions (labeled correct in CodeContests) under the same input, following the existing work~\cite{li-science22}.
Any input leading to inconsistent outputs will be filtered out (detailed in Appendix~\ref{sec:validator}).
After these, we rank the tests for each solution in the dataset based on the slowdown they introduce. 
We then select the top ten longest-running tests for each program and aggregate them as part of our benchmark, \bench.
\subsection{Main Results}
\label{sec:quantitative_results}

To evaluate the effectiveness of \bench tests in stressing performance, we compare the slowdown \bench brings to the programs against those by \evalperf, \tgprompt, and \perffuzz.

% As a reminder \bench includes the union of top ten tests that slow each program the most.
% For a fair comparison, for each baseline we collect the a set of tests following the same process of taking the union of top ten tests that determine a program to execute most CPU instructions.

\begin{table*}[!t]

\centering
\setlength{\tabcolsep}{12pt}
\renewcommand{\arraystretch}{1}
\small

\caption{\tool versus baselines (described in Section~\ref{sec:setup}) and its ablation.}

\begin{tabular}{lllrrr}
\toprule
\multirow{2}{*}{Technique} & \multicolumn{2}{c}{\# of instructions ($\times10^8$)} & \multirow{2}{*}{Win rate} & \multicolumn{2}{c}{Slowdown over CC} \\
              & \multicolumn{1}{c}{Average} & \multicolumn{1}{c}{Median} & & \multicolumn{1}{r}{Average} & \multicolumn{1}{r}{Median} \\
\midrule
\rowcolor[gray]{0.95}\multicolumn{6}{c}{\bf Compare to baselines} \\
\tool & \textbf{5.96} & \textbf{0.75} & \textbf{60\%} & \textbf{363$\times$} & \textbf{1.65$\times$} \\
\tgprompt & 3.87 (\textcolor{red}{$\downarrow$}1.5$\times$) & 0.60 (\textcolor{red}{$\downarrow$}1.3$\times$) & 12\% & 275$\times$ & 1.52$\times$ \\
\perffuzz & 3.29 (\textcolor{red}{$\downarrow$}1.8$\times$) & 0.43 (\textcolor{red}{$\downarrow$}1.7$\times$) & 11\% & 149$\times$ & 1.61$\times$ \\
\evalperfslow & 3.23 (\textcolor{red}{$\downarrow$}1.8$\times$) & 0.44 (\textcolor{red}{$\downarrow$}1.7$\times$) & 8\% & 146$\times$  & 1.63$\times$ \\
\evalperfrandom & 3.21 (\textcolor{red}{$\downarrow$}1.9$\times$) & 0.45 (\textcolor{red}{$\downarrow$}1.7$\times$) & 9\% & 166$\times$ & 1.54$\times$ \\
\midrule
\rowcolor[gray]{0.95}\multicolumn{6}{c}{\bf Ablations} \\
\tool & \textbf{5.96} & \textbf{0.75} & \textbf{65\%} & \textbf{363$\times$} & \textbf{1.65$\times$} \\
\wedgenoinstrument & 4.02 (\textcolor{red}{$\downarrow$}1.5$\times$) & 0.21 (\textcolor{red}{$\downarrow$}3.6$\times$) & 29\% & 159$\times$ & 1.13$\times$ \\
\wedgedefaultmutator & 1.54 (\textcolor{red}{$\downarrow$}3.9$\times$) & 0.01 (\textcolor{red}{$\downarrow$}>75$\times$) & 4\% & 12$\times$ & 0.99$\times$ \\
\wedgeafl & 1.49 (\textcolor{red}{$\downarrow$}4.0$\times$) & 0.01 (\textcolor{red}{$\downarrow$}>75$\times$) & 2\% & 30$\times$ & 0.99$\times$ \\
\bottomrule
\end{tabular}

\label{tab:efficiency-main}

\end{table*}

\begin{figure*}[!t]

\centering

\subfloat[\tool vs. \tgprompt]{
        \includegraphics[width=0.31\linewidth]{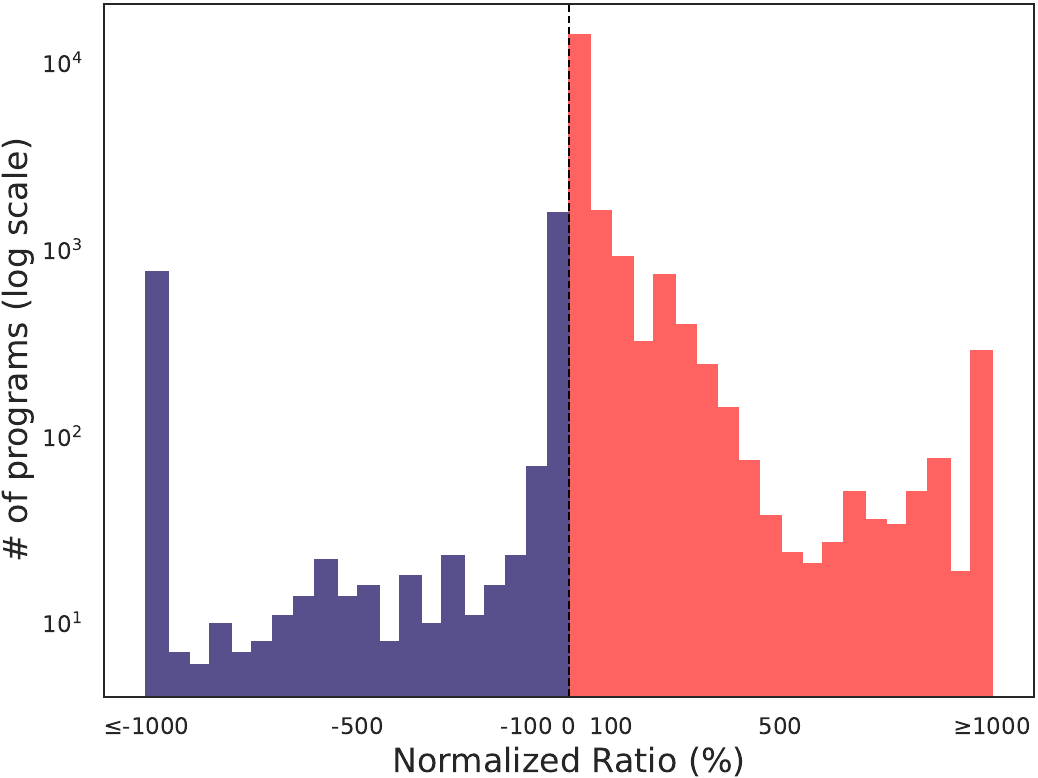}
        \label{subfig:histogram-wedge-vs-tg}
    }
\subfloat[\tool vs. \evalperfslow]{
        \includegraphics[width=0.31\linewidth]{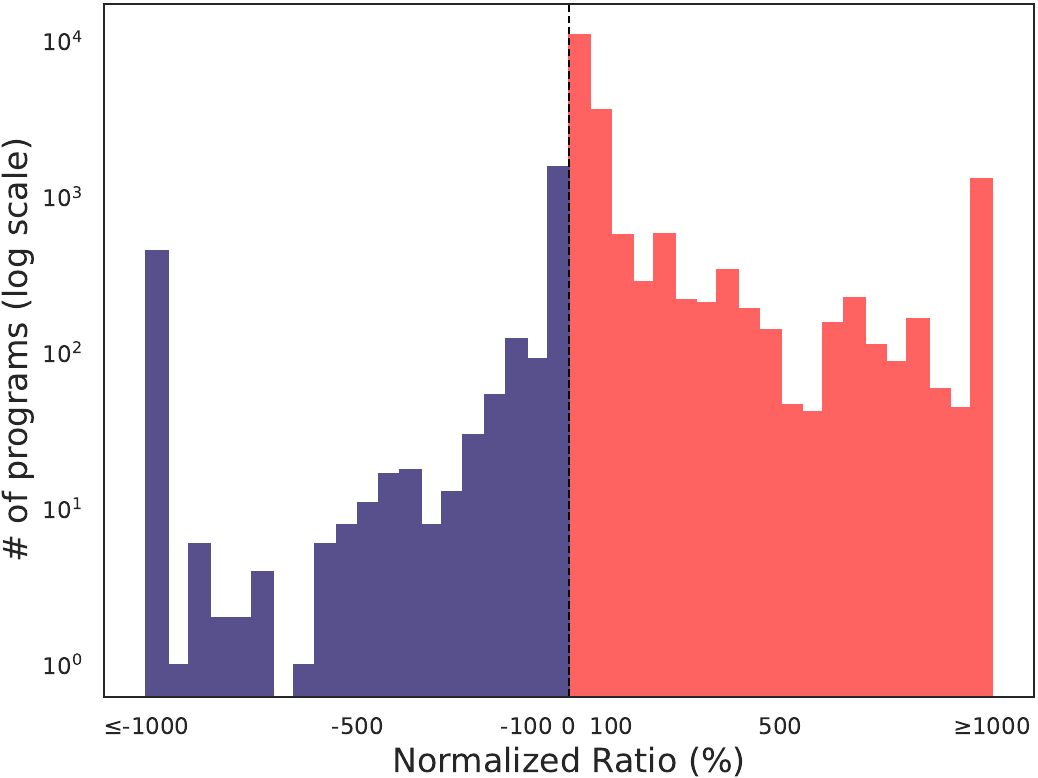}
        \label{subfig:histogram-wedge-vs-evalperf-slow}
    }
\subfloat[\tool vs. \perffuzz]{
        \includegraphics[width=0.31\linewidth]{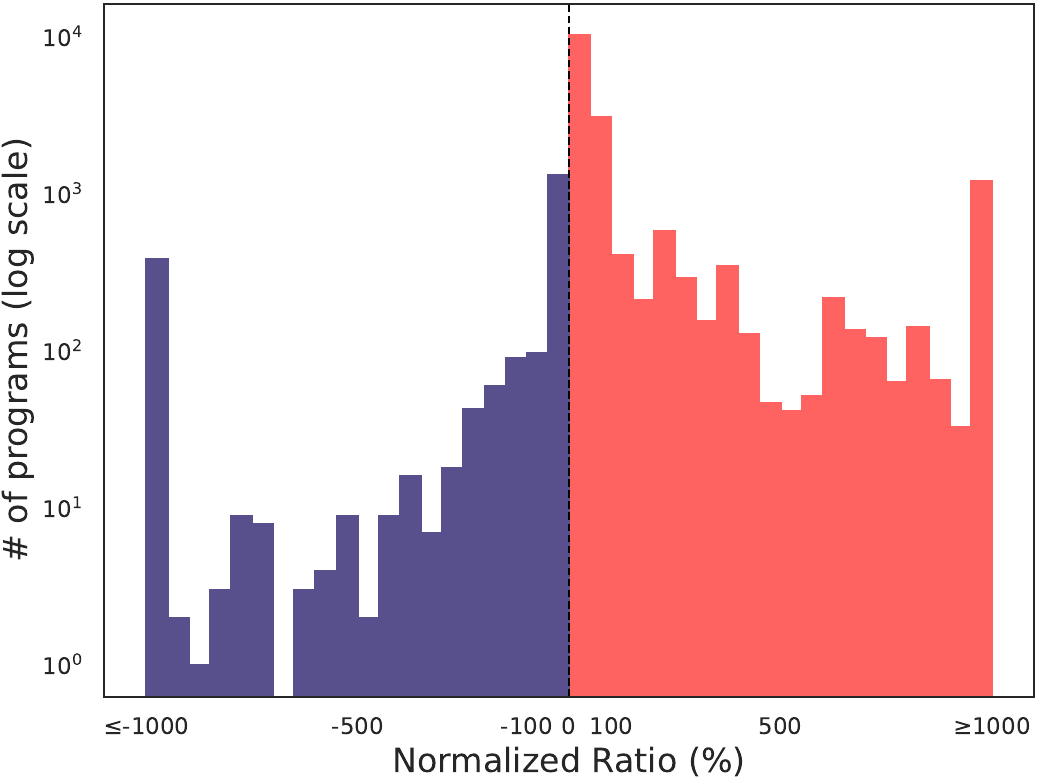}
        \label{subfig:histogram-wedge-vs-perffuzz}
    }
    
\caption{A head-to-head comparison between \bench (\textcolor{CustomOrange}{\ding{110}}) and the baseline tests (\textcolor{CustomPurple}{\ding{110}}). The bars represent the number of programs where one incurs a larger number of CPU instructions. x-axis shows the corresponding ratio between the corresponding CPU instruction counts. Since the two \evalperf variants show similar distributions, we only include  \evalperfslow here (see Section~\ref{appx:head-to-head}).
}
\label{fig:histogram-head-to-head}
\end{figure*}

Table~\ref{tab:efficiency-main} shows that tests generated by \tool lead programs to execute, on average, 84.5\% and 85.7\% (70.5\% and 66.7\% median) more CPU instructions than the two variants of \evalperf, respectively.
They also have 54\% (25\% median) more CPU instructions than \tgprompt.
\bench tests dominate the number of programs (59\%) where they run the slowest among all the other baselines (win rate).
\cref{fig:histogram-head-to-head} visualizes the win rate by running head-to-head comparison between \tool and the baselines.
% Appendix~Section~\ref{sec:appendix-running-time} shows similar trends for physical execution time.
We also compute the slowdown that the tests achieved over the default tests in \codecontest.
On average, \bench tests outperform \evalperf ones by 2.3$\times$ and \tgprompt by 1.3$\times$.
\cref{fig:histogram-head-to-head} illustrates a head-to-head comparison between \bench and the baselines, where \bench's tests slows significantly more programs compared to the other baselines.
Analyzing physical running times reveals similar trends (see Appendix~\ref{appx:measurements}).
% For each program, it shows which of the two techniques' generated tests (both top-10 slowest inputs) executes more instruction.
% A positive percentage indicates that \bench tests determine a program to execute more instructions a program, while a negative percentage indicates that the other technique in the head-to-head comparison does.
% We then plot the corresponding histogram with 10\%-size bins.
% The right half indicates programs that running \bench tests slow down more than the baseline (\textcolor{CustomOrange}{\ding{110}}), and the left half shows the reverse (\textcolor{CustomPurple}{\ding{110}}).

We extensively analyzed the \pccsmall as well as the test generators synthesized by \tool and the other baselines and benchmarks.
We observe that the inputs generated by \tool focus more on the inefficient implementation in the code identified by the performance-characterizing constraints, while those by \evalperf are optimized to stress the input length specified in the problem statement.
\tgprompt, while not explicitly implemented to maximize bounds, faced challenges in reasoning about holistic program behaviors end-to-end.
Even with chain-of-thought prompting, it still reduces to mostly generic \lengthstressing inputs specific to the problem statement (e.g., large graphs for graph-based problems).
We leave the detailed description of these qualitative studies in \cref{appx:case-studies} due to space constraints.

Similarly, while \perffuzz is designed to trigger worst-case behavior by favoring inputs that execute more control-flow graph's edges, we observe that \perffuzz still ends up generating length-stressing inputs.
% Without explicitly identifying and exercising the performance-characterizing constraints, \perffuzz tends to overlook inputs that are stressing but not captured by the existing branches.
Without explicitly identifying and exercising the performance-characterizing constraints, \perffuzz can overlook inefficient implementations when it is guided only by the coverage of existing branches in the code and the profiler.
This is because the existing branches in the code are often irrelevant to the efficient behavior, while the profiler may not capture the root cause of performance bottleneck, as the input generated so far may not have exercised the inefficient implementation yet.

\para{Ablations.} 
We ablate the two designs related to \pcc: (1) guiding the mutator generation with constraints and (2) instrumenting the program with constraint checker code.
% For (1), we consider a constraint-agnostic mutator as the baseline, generated by prompting without \pcc, following the approach similar to those in generating TG-prompt (Section~\ref{sec:setup}).
For (1), we consider the AFL++ mutator as the baseline.
For (2), we consider the original program in the baseline.

Table~\ref{tab:efficiency-main} shows that \tool's generated tests are on average 48.3\% and 128.3\% slower (in terms of CPU instruction count and relative slowdown) than \wedgenoinstrument, showing that the instrumented programs with constraint checkers can effectively guide fuzzing.
Similarly, \tool's generated tests incur 287\% more CPU instructions than those generated by \wedgedefaultmutator (with default mutator).
On 63.02\% solutions, \tool tests are slower than \wedgenoinstrument tests (with a significance value of 0.05, base on Mann-Whitney test~\cite{mcknight2010mann}).
% This underscores the importance of the constraint-aware mutator.

\subsection{Utility of \bench}
\label{sec:tests-utility}

% Execution feedback has proven valuable for LLM-based code 
%     edits~\cite{chunqiu-issta24,peng2024perfcodegen,huang-neurips24}.
% However, its effectiveness hinges on tests that accurately expose 
%     performance bottlenecks.
As described in Section~\ref{sec:setup}, we investigate the utility of \bench by comparing \bench tests to the default \codecontest tests (\publicfiveprofile) that only evaluate the correctness on (1) improving LLM-based code optimizations (\effilearner~\cite{huang-neurips24}) based on the execution feedback, and (2) fairly measuring performance improvement where the baseline's evaluation relied only on correctness tests (\pie~\cite{shypula-iclr24}).
To ensure a fair comparison, we adopt the exact same evaluation setup and metrics used by the two baselines.
For example, we include memory usage to evaluate how \bench improves \effilearner.
Since \effilearner relies on the original CodeContests tests, yet for about 15\% problems have less than ten tests available, we instead use top-5 slowest tests per solution. 

\paragraph{Improving code optimization with execution feedback.}
% \label{sec:effilearner}
We collect a corpus of 280 slow Python solutions from 56 problems in \bench following the \effilearner's filtering strategy.
For each solution, 
% we optimize it using \effilearner based on its original tests and compare with that using \bench tests. 
% We profile each run and record physical time and memory utilization statistics.
% We then annotate the top bottleneck functions with these metrics into the tool's original prompt format.
we run \effilearner with three different prompts to let \effilearner optimize the code.
(1) we use the solution code alone with no execution feedback (\noprofile). 
(2) we use code annotated with profiling information from running the original \codecontest default tests (\publicfiveprofile).
(3) we use code annotated with profiling information derived from our \bench tests (\wedgeprofile).
We consider both OpenAI GPT-4o and DeepSeek V3 as the backends for \effilearner.

\begin{table*}[!t]
    \centering
    \setlength{\tabcolsep}{7pt}
    \renewcommand{\arraystretch}{1.1}
    \small

    \caption{Running \effilearner for code optimization using execution feedback from different types of test sets. \bench improves \effilearner the most.}
    \label{tab:effilearner-experiment}
    % \begin{tabular}{rlrrrr}
    %     \toprule
    %     Model &  & \makecell{Execution \\ time (s)} &
    %     \makecell{Total memory \\ usage (Mb * s)} &
    %     \makecell{Max memory \\ usage (Mb)} &
    %     \makecell{Execution \\ time (\#inst)} \\
    %     \midrule
    %     \multirow{3}{*}{GPT-4o} 
    %       & \noprofile     & 15.66\%              & 21.70\%         & 2.78\%           & 31.39\% \\
    %       & \publicfiveprofile & 22.57\%              & 27.78\%         & \textbf{11.89\%} & 39.89\% \\
    %       & \wedgeprofile  & \textbf{26.47\%}     & \textbf{35.10\%}& 11.82\%          & \textbf{49.31\%} \\ 
    %     \midrule
    %     \multirow{3}{*}{DeepSeek-V3} 
    %       & \noprofile     & 6.77\%               & 12.49\%         & 2.30\%           & 20.48\% \\
    %       & \publicfiveprofile & 12.29\%              & 20.43\%         & 1.93\%           & 35.01\% \\
    %       & \wedgeprofile  & \textbf{18.99\%}     & \textbf{23.75\%}& \textbf{9.21\%}  & \textbf{40.26\%} \\ 
    %     \bottomrule
    % \end{tabular}
    \begin{tabular}{lrrrrrrrr}
    \toprule
    \multirow{2}{*}{\makecell{Test set}} 
    & \multicolumn{2}{c}{\makecell{Execution time (s)}} 
    & \multicolumn{2}{c}{\makecell{Total memory \\ usage (Mb * s)}} 
    & \multicolumn{2}{c}{\makecell{Max memory \\ usage (Mb)}} 
    & \multicolumn{2}{c}{\makecell{CPU instructions}} \\
    & GPT-4o & DS-V3 & GPT-4o & DS-V3 & GPT-4o & DS-V3 & GPT-4o & DS-V3 \\
    \midrule
    \noprofile     
    & 15.66\% & 6.77\% 
    & 21.70\% & 12.49\% 
    & 2.78\%  & 2.30\%  
    & 31.39\% & 20.48\% \\
    
    \publicfiveprofile 
    & 22.57\% & 12.29\% 
    & 27.78\% & 20.43\% 
    & \textbf{11.89\%} & 1.93\%  
    & 39.89\% & 35.01\% \\
    
    \wedgeprofile  
    & \textbf{26.47\%} & \textbf{18.99\%} 
    & \textbf{35.10\%} & \textbf{23.75\%} 
    & 11.82\% & \textbf{9.21\%} 
    & \textbf{49.31\%} & \textbf{40.26\%} \\
    \bottomrule
\end{tabular}
\end{table*}

Table~\ref{tab:effilearner-experiment} illustrates that \bench tests achieves the best performance improvement.
\effilearner can optimize the code to execute 24\% less instructions (or approximately 10 percentage points), run 17\% faster, and use 25\% less memory on average when providing GPT-4o with \bench execution profiles as opposed to their original setup.
Similarly, \effilearner can optimize the code to execute 15\% less instructions, run 46\% faster, and use 16\% less memory when providing DeepSeek with the same \bench-driven execution profiles.

\paragraph{Evaluating code optimization fairly.}
% \label{sec:pie}
We show how \bench can measure performance improvement claimed by existing code optimization more fairly than the correctness test.
% , in addition to LM-backed feedback-driven code optimizations.
To this end, we consider \pie~\cite{shypula-iclr24}, a state-of-the-art \llm-based code optimization based on finetuning, but relied on the default correctness tests to measure their performance improvement.
We select their three most effective models (CodeLlama 13b) finetuned with the following different datasets: 
(1) HQ (high-quality) data annotated by the authors (\piehq); 
(2) performance-conditioned data to optimize C++ programs annotated with a target optimization score reflecting its potential ``peak performance'' (\piecond);
(3) all data from the entire \pie dataset.
We then adapt our program selection to match the requirements of \pie (details in~\cref{appx:dataset-filtering})

We follow the same set of metrics as~\cite{shypula-iclr24} by measuring the average relative speedup between the original and optimized code in instruction counts and physical time, as well as the percentage of programs that the \llm models can optimize by at least 10\% (\%Opt)~\cite{shypula-iclr24}.
\cref{tab:pie-experiment} illustrates how our tests better characterize the performance bottlenecks.
\bench outperforms the \codecontest default tests (\publicfiveprofile) and its top five slowest tests (\defaultslowfive) by 24\% to 149\% in terms of instruction counts and by 5\% to 27\% in terms of physical time.
It also helps discover that between 7\% and 48\% more programs have actually been meaningfully optimized and run at least 10\% faster.

\begin{table*}[!t]

\centering
\renewcommand{\arraystretch}{1.1}
\small

\caption{Pie Experiment: average speedup and fraction of optimized programs, i.e., at least 10\% faster (\%Opt) evaluated by different test sets following \cite{shypula-iclr24}. The top-performing test set is highlighted.}
\label{tab:pie-experiment}
\setlength{\tabcolsep}{3.5pt}
\begin{tabular}{lrrrrrrrrrrrr}
\toprule
\multirow{2}{*}{\makecell{Test set}} 
& \multicolumn{3}{c}{Speedup (\#inst)} 
& \multicolumn{3}{c}{Speedup (time)} 
& \multicolumn{3}{c}{\%Opt (\#inst)} 
& \multicolumn{3}{c}{\%Opt (time)} \\
& \piehq & \piecond & \pieuncond 
& \piehq & \piecond & \pieuncond 
& \piehq & \piecond & \pieuncond 
& \piehq & \piecond & \pieuncond \\
\midrule
\defaultall     
& 1.32 & 1.98 & 1.21 
& 0.93 & 1.13 & 1.32 
& 6.0\% & 16.2\% & 4.4\% 
& 16.6\% & 29.2\% & 56.5\% \\

\defaultslowfive 
& 1.30 & 1.85 & 1.21 
& 0.94 & 1.13 & 1.32 
& 5.5\% & 15.6\% & 4.4\% 
& 18.1\% & 27.9\% & 56.5\% \\

\wedgeslowfive  
& \textbf{1.62} & \textbf{1.99} & \textbf{3.01} 
& \textbf{1.18} & \textbf{1.29} & \textbf{1.38} 
& \textbf{12.1\%} & \textbf{18.8\%} & \textbf{30.4\%} 
& \textbf{26.6\%} & \textbf{41.6\%} & \textbf{60.9\%} \\
\bottomrule
\end{tabular}

\end{table*}

\subsection{Sensitivity Analysis}
\label{sec:sensitivity_analysis}

% To understand the impact of our design
%     choice, we ablate the two key decisions: 
%     (1) involving pc-constraints
%     in mutator generation, 
%     (2) instrumenting the original 
%     program with \textit{constraint checker} code.
% For (1), we use a \textit{constraint-agnostic} 
%     mutator for comparison.
% The constraint-aware mutator and 
%     constraint-agnostic mutator are generated by prompting with and without the 
%     pc-constraints.
% For (2), we use the original program for comparison.

% Table~\ref{tab:efficiency-main} shows the performance of \tool and ablated versions.
% \tool tests (using top-10 inputs) are on average 63.03\% and 62.34\% slower (in terms of relative slowdown and \#instructions) than \wedgenoinstrument.
% The results show using instrumented programs (programs with constraint checkers inserted) can provide guidance to fuzzing, thus generating slower tests compared with not inserting the checkers.
% Similarly, \wedgedefaultmutator tests are on average 155.17\% and 156.99\% slower than \wedgeafl.
% On 63.02\% solutions \wedgenoinstrument tests are slower (with a significance value of 0.05, according to Mann-Whitney test~\cite{mcknight2010mann}).
% This underscores the importance of the performance-constraint-aware mutator.

\para{Discriminative power of performance-characterizing constraints.}
To investigate whether and how \tool-generated performance-characterizing constraints can indeed capture performance-stressing inputs, we select 810 programs in CodeContest where both constraint-satisfying and constraint-violating inputs exist.
Results show that constraint-satisfying inputs are, on average, 38.6$\times$ slower than constraint-violating inputs.
We conduct a Mann-Whitney test~\cite{mcknight2010mann}, and constraint-satisfying inputs are significantly slower (with a significance value $p < 0.05$) than constraint-violating inputs on 92.84\% programs.
% It demonstrates the performance-characterizing constraints generated by \tool are discriminative across most programs.

\para{Impact of constraints in guiding fuzzing.}
To better understand the impact of guidance of constraints (including mutator and code instrumentation for coverage guidance), we calculate the ratio of constraint-satisfying inputs (out of valid inputs) per strategy.
Result shows that the ratios of constraint-satisfying inputs among generated inputs of \wedgeafl, \wedgedefaultmutator, \wedgenoinstrument, and \tool are 40.42\%, 41.44\%, 77.62\%, and 80.48\%, respectively.
In other words, both involving performance-characterizing constraints and constraint checker code contribute positively to the ratio of constraint-satisfying inputs.
Furthermore, strategies that yield a higher proportion of constraint-satisfying inputs tend to achieve better performance (see Section~\ref{sec:quantitative_results}), indicating that satisfying performance-characterizing constraints correlates with the generation of more stressing test inputs.

\para{Effect of input size.} 
We investigate how input size affects the effectiveness of \bench considering that leveraging fuzzing to generate large inputs is a known challenging problem~\cite{lemieux_issta2018}.
We observe our framework outperforms the baselines by larger margins when we further restrict the input size to be less than 1KB. 
In particular, for problems whose inputs are less than 1KB, the slowdown achieved by \tool is $3\times$, almost double that on the entire problems without such restrictions, 1.5$\times$.
These findings underscore that the performance-stressing characteristics of our tests stem from inputs being designed to target implementation-specific bottlenecks rather than being simply length-stressing.
We put the detailed results in \cref{appx:input-size-effect} due to space constraints.
\section{Related Work}
\label{sec:rel-work}

\para{Performance fuzzing}. A popular line of related work aims to trigger performance bugs by automatically crafting worst-case inputs~\cite{pestios_ccs2017,lemieux_issta2018,zhang_ase2011,burnim_icse09,chen_icse16,padhye-oopsla19}.
For example, SlowFuzz~\cite{pestios_ccs2017} and PerfFuzz~\cite{lemieux_issta2018} are feedback-driven fuzzers that search for inputs causing extremely long execution cost. 
% PerfFuzz in particular uses multi-dimensional feedback to independently maximize execution counts for many program points, thus finding a variety of inputs that exercise different frequently executed (hot) paths.
FuzzFactory~\cite{padhye-oopsla19} generalizes this idea by allowing developers to define custom performance-oriented feedback metrics and integrate them into a fuzzing framework. 
These tools have been effective at uncovering inefficiencies related to algorithmic complexity or poor resource utilization. 
However, they rely on runtime instrumentation or heuristics (e.g., counters for loops or allocations) as well as specific performance hints or signals, often manually crafted, to guide the input search. 
\tool complements this line of research by automating the synthesis of diverse performance-characterizing constraints as the test oracle.

\para{Performance bug detection}.
% Researchers have developed numerous tools to 
Numerous tools have been invented to detect performance bugs by identifying inefficient code patterns, costly loops, repeated computations, and suboptimal usage of data structures~\cite{aguilera03, jin-pldi12, nistor-icse13, nistor-icse15, olivo-pldi15, xu-pldi09, xu-pldi10, dufour-issta07, jli18}.
While \tool could be extended to detect performance bugs, it focuses more on translating the performance-stressing symptoms into fuzzer-amenable constraints.
Therefore, \tool focuses more on evaluating and improving existing LLM-based code optimization approaches, as opposed to finding performance bugs in large-scale systems.
    
\para{LLM-assisted test generation}. 
% A growing body of research focuses on LLM-assisted test generation~\cite{zhang2024ecg, hu2023augmenting, zhang2024llamafuzz, yang2023kernelgpt, yang2024whitefox, deng2024large, deng2024large, xia2024fuzz4all, liu2024oss-fuzz-gen, shi2024harnessing, zhang2024effective, lemieux-icse23, zhu2025locus}.
% patch_2
A growing body of research focuses on LLM-assisted test generation~\cite{zhang2024ecg, hu2023augmenting, zhang2024llamafuzz, yang2023kernelgpt, yang2024whitefox, deng2024large, xia2024fuzz4all, liu2024oss-fuzz-gen, shi2024harnessing, zhang2024effective, lemieux-icse23, zhu2025locus}.
% For instance, CodaMosa~\cite{lemieux-icse23} augments search-based test generation by directly generating inputs (test cases) to escape coverage plateaus.
% CodaMosa~\cite{lemieux-icse23} augments search-based input fuzzing with LLMs to escape coverage plateaus.
% It runs an off-the-self fuzzer until coverage no longer increases, identifies call statements that rarely executes, prompts an LLM to generate inputs that attempt to cover them, and adds these to the fuzzer's input corpus.
% Libro~\cite{kang-icse23} iteratively prompts LLMs to generate and rank tests based on bug specifications.
% TitanFuzz~\cite{deng-issta23} uses LLMs to directly perform input generation and mutation to fuzz deep learning libraries.
% In contrast, TitanFuzz~\cite{deng-issta23} uses LLMs as zero-shot fuzzers by having two collaborating agents perform generation- and mutation-based fuzzing.
% \evalperf uses LLMs to synthesize the input generator, which is controlled by a ``scale'' parameter, leading the generator to focus mainly on length-stressing inputs.
% The formulation adopted in these techniques can pose a high reasoning burden on LLMs, hindering reliable reasoning about program behavior.
% 
Most of these works leverage LLMs to directly generate the stressing inputs or the input generators based on the code under test.
This usually requires tracking and reasoning over long-range control and data dependencies in the program, posing significant reasoning burdens to LLMs to infer the desired inputs all the way from specific program points deep in the code.
In contrast, \tool alleviates LLMs' role from end-to-end input synthesis to only generating local performance-characterizing predicates, mitigating the burden introduced by the potentially long-context reasoning.

\section{Discussion and Conclusion}
\label{sec:conclusion}

\para{Limitations.}
The limitations of \tool are threefold.
First, \tool incorporates prompting and fuzzing for each solution, thus incurring considerable token cost and execution overhead.
Second, performance-characterizing constraint reasoning depends on mining high-quality contrastive input pairs, thus requiring either an existing test input corpus and/or additional executions.
Third, the underlying fuzzer often suffers from the input length constraints.
For example, AFL++~\cite{fioraldi-woot20} can only mutate input data smaller than a threshold, e.g., 1MB by default (and we modified it to 10MB), thus cannot handle arbitrarily large inputs.
% If a constraint is complex and deeply nested, the fuzzing engine might not be strong enough to satisfy the constraint within the given time budget.

\para{Future work.}
% We develops a methodology to generate stress tests by combining the advantage of reasoning ability of LLMs and searching ability of fuzzing, which can inspire future research in performance test generation (on real-world software).
Our key insight is to decompose the problem of stress test generation into performance-characterizing constraint reasoning and constraint-guided search, where the former can benefit from LLM's code reasoning capability and the latter can leverage efficient input search tools based on fuzzing.
% which can inspire future research in performance test generation (on real-world software).
% The similar spirit generalizes to various performance and security debugging tasks.
Future work includes extending \tool to real-world projects and generating test oracles beyond performance evaluation.
% First, \tool's reliance on fuzzing that readily supports large projects.
% And its design in constraining LLMs for local performance constraint reasoning makes it promising to generalize to larger projects. 
% Specifically, \tool can be applied to generate performance constraints at the local function of the project.
% Such a local reasoning avoids relying on LLM's limited capabilities in long-context reasoning~\cite{liu2023repobench,jimenez2023swe,shetty2025gsochallengingsoftwareoptimization,he2025swe}, while fuzzers bridge such gaps to enable global input search for both whole-project or library testing~\cite{ossfuzz}. 
% In addition, optimizing at the local code context can still benefit the entire project~\cite{Caballar2023Scalene,bala2000dynamo,sasnauskas2017souper}.

\paragraph{Conclusion.}
We introduced \tool, a framework to evaluate and improve code efficiency by generating performance-characterizing constraints with LLMs and guided fuzzing to explore performance-stressing inputs.
% We transform the problem of stress test input generation into a search problem, and our generated constraints effectively help divide the search space and guide the search process.
% \tool can outperform all baseline techniques on 59\% solutions, on average 1.5$\times $better than the second-best baseline.
% The generated constraints are strongly discriminative and effective in guiding fuzzing, as the constraint-satisfying inputs are significantly slower than the constraint-violating inputs, by 51.46$\times$ on average, and the fuzzer guided by constraints generates slower inputs than no constraint guidance on 93.73\% programs.
We released our performance-stressing tests, along with the CodeContest programs, as a new benchmark \bench at \website.
With \bench tests, we have demonstrated that \tool helped better evaluate and substantially improve existing code optimization techniques.

\section*{Acknowledgement}

We thank all the anonymous reviewers, Weichen Li, and Jiawei Liu for their constructive and insightful comments and feedback, which significantly improved this paper.
This work was supported in part by the OpenAI Research Access Program~\cite{openairesearchprogram}.
% This work was supported by NSF grant CCF-2119184 and CNS-1764039.
Results presented in this paper were obtained using the Chameleon testbed~\cite{chameleoncloud} supported by the National Science Foundation.

\bibliographystyle{unsrtnat}
\bibliography{references}

\newpage

\appendix

\section{Detailed Design Decisions}
\label{appx:design-decisions}

\subsection{Filtering Policies}
\label{appx:dataset-filtering}

\para{Filtering process for evaluating \tool.} The original CodeContest dataset (``train'' split, ``\codeforces'' section~\cite{codecontests-p633A}) contains over 3,000 problems with hundreds C++ solutions each.
It is infeasible to evaluate that many executions due to both computational and monetary costs.
To address this, we focus on a smaller yet meaningful subset by applying the following filtering criteria.

\noindent (1) \textit{Sufficient computation.} We remove problems whose solutions never exceed 100,000 instructions on any input to reduce the impact of potential noise following \cite{liu-colm24}.

\noindent (2) \textit{Sufficient solutions.} We remove problems with fewer than 10 correct solutions because we need to run fuzzing on a reasonable amount of correct solutions to obtain meaningful measurements.

\noindent (3) \textit{Sufficient test inputs.} We remove problems with fewer than five default tests because we need enough inputs for fuzzing to be effective and also to identify contrasting input pairs (~\cref{sec:exec-feedback-collection}).

\noindent (4) \textit{Single output.} We filter out problems that accept multiple correct outputs.

\noindent (5) \textit{Diversified performance.} We only include problems with diversified performance across the solutions. 
Specifically, we use the Coefficient of Variance (CV) to measure the diversity, following~\cite{liu-colm24}. 
Low diversity means most solutions have similar performance, indicating the solutions are likely optimal or close to optimal.
Intuitively, there are fewer opportunities to identify performance-characterizing constraints on the optimal solutions.

\para{Additional filtering to accommodate \effilearner.} 
Note that the original \effilearner\xspace explores using 
    profiling information to improve \llm-generated code, while 
    we adapt it to our scenario, \ie improving slow user 
    solutions.
To this end, we extract a subset of problems and solutions from 
    the 300 problems in \bench. Specifically, in addition to the 
    \textit{sufficient computation} criterion above, we applied 
    the following criteria:
% 1) Python solutions with function definitions. Because the profilers \effilearner\xspace used work on Python function level code.

% 2) Slow enough, containing at least 100K instructions on average and at least twice as slow as the fastest solution for the same problem. This can ensure the solution code has some complexity and there's considerable space for optimization.

% 3) Long enough, containing at least 50 lines. Because there might be fewer places that can potentially get optimized in smaller programs.

% We select problems where at least 5 such solutions exists and select the 5 slowest solutions for each problem to consist our evaluation dataset here. In the end, we got 56 problems from \bench.

% We apply the following criteria to select the set of solutions to evaluate how \tool tests can boost feedback-based code optimization:
\noindent (1) \textit{Python code.} Instead of C++, we focus 
    on Python solutions as \effilearner is designed for Python. 
\effilearner also requires these Python programs to 
    contain function definition, \ie with \codeIn{def} keyword, as the profiling tools work at the function level.
 
\noindent (2) \textit{Room for code optimization.} Solutions 
    should be relatively slow, namely suboptimal, so that there's 
    space for improvement.
Thus, we include only solutions that execute approximately 
    twice the number of instructions when compared to the 
    fastest (correct) solution for that particular problem.
 
Similar to our main filtering criteria above, we select problems that have at least five such solutions. 
We then select the five slowest ones per problem to form our evaluation dataset. 
Ultimately, we end up with 56 problems and 280 Python solutions.

\para{Additional filtering to accommodate \pie.}
The original \pie framework~\cite{shypula-iclr24} was evaluated on solutions to coding problems from the CodeNet dataset, a subset of CodeContests.
Following the authors' original experimental setup, we focus on C++ solutions and, in addition to the \textit{sufficient computation} and \textit{room from improvement} criteria above, we also require solutions to be relatively short ($\leq$50 LoC) to accommodate the small \llm context window that \pie requires.

\subsection{Iterative Mutator Generation}
\label{sec:mutator}
Although \afl~\cite{fioraldi-woot20} has shown superior performance in terms of coverage-guided test input generation, it is not ready to use off the shelf.
The default mutators of \afl engine are optimized for compact data formats - say, images, multimedia, compressed data, \etc supporting operations like flipping bits, inserting bytes, changing bytes, \etc not aware of the validity constraints of inputs.
Therefore, such byte-level mutators will produce many invalid inputs that violate the input constraints of the coding problems.
Instead of using the default mutators, we rely on the \textit{custom mutators} interface of \afl, which allows the user to specify a customized Python or C++ mutator script as the mutator engine. 
We prompt the model with problem description $\mathcal{S}$, summarized constraints $\mathbb{C}_\mathcal{P}$, \afl provided mutator example, etc.

Different from \evalperf~\cite{liu-colm24} and \tgprompt~\cite{huang-neurips24,peng2025coffecodeefficiencybenchmark,du2024mercury} style \textit{generator synthesis}, \textit{mutator synthesis} is more challenging since in addition to making sure the mutated input follows the input constraints, it also needs to be robust enough to mutate inputs in various shapes.
To this end, we propose to use an iterative mutator generation approach.
The intuition comes from recent works~\cite{chunqiu-issta24} showing LLMs are good at fixing programs conversationally.
For each generated mutator, we launch a dry run for three minutes. 
If the dry run successfully exited with a number of new test inputs generated, the mutator is labeled as ``pass''.
Otherwise, we append the failing message such as ``IndexError: list assignment index out of range'' to the conversation and ask the \llm to fix the issue.
The prompting-and-dry-run loop terminates when a good mutator is produced, or the maximum number of rounds (ten in our experiments) have been tried.
The evaluation shows only the ratio of successful mutators of single round generation is only \numtodo{80.13\%}, while it's improved \numtodo{82.96\%} after one round and \numtodo{83.35} after ten rounds, demonstrating the effectiveness of iterative refining.

\subsection{Input Validation with Consistency Checks}
\label{sec:validator}

\para{LLM-based input validator.}
Both fuzzing and prompting-based generator generation can produce a large amount of test inputs. 
However, the generated inputs might violate the input format or specifications in the problem description.
To tackle this problem, \citet{liu-neurips23} adopted a \textit{programming by contract} philosophy by \textit{manually} annotating function pre-conditions in the form of code assertions (\eg \codeIn{assert n > 0}) to ensure the test inputs for the function are well-formed. 
However, manual annotation is known to be error-prone and expensive~\cite{brants-emnlp07}.
On the other hand, synthesizing a validator is generally challenging as there's no ground truth.

We rely on the rich test cases (which are labeled as correct by the open-judge platform) in the dataset to reduce wrong validators, following the paradigm of Programming By Example (PBE), \ie a good validator should not label any valid inputs as invalid.
Specifically, we iteratively prompt the \llm to generate a validator Python script and execute it on the ``public'' and ``private'' test cases.
If some tests fail, we append the failing message to the conversation and ask the model to fix the validator, until all tests pass or it reaches the retry limits.
If no good validator is generated for a problem, it is excluded from our evaluation set.
In the end, we can successfully generate \numtodo{289} validators within five rounds.
Among them, \numtodo{280} validators are generated successfully in the first round.
Note that this approach does not ensure the correctness of the validators as the synthesis problem is generally undecidable, but it gives us higher confidence on the reliability of the validators.

\para{Consistency check.}
In addition to the generated validator, we introduce a  
    ``consensus'' consistency check to further filter 
    out invalid inputs.
For each generated input, we execute all correct solutions under the input and check whether 95\% of them are consistent.
Inputs leading to over 5\% inconsistent results will be discarded.
The intuition is that a well-formed input should be processed correctly by all correct solutions.
We apply the validators and consistency check to all techniques to ensure a fair comparison.

\subsection{Implementation and System Environment} 
\label{appx:settings}
We run experiments on six x86-64 machines equipped with a 24-core Intel Xeon Gold 6126 CPU with 192GB of RAM.
Each machine runs Ubuntu 20.04 LTS (kernel version 5.4.0).
We use the OpenAI GPT-4o (gpt-4o-2024-08-06) and DeepSeek V3 (deepseek-v3-2024-12-26) with a temperature of 0.8 and max\_length of 4,096.
We use GPT-4o as the backbone LLM of \tool and all baseline techniques.
For \effilearner utility experiment~\ref{sec:tests-utility}, we use both GPT-4o and DeepSeek V3 for prompting.
For \pie utility experiment~\ref{sec:tests-utility}, we use the same settings (temperature of 0.8, max\_length of 4,096) for the three evaluated fine-tuned models.

In the contrasting-input-pair mining stage, \tool will select no more than 10 solutions where at least one contrasting input pair exists.
It then reasons about and generates performance-characterizing constraints per solution.
Each constraint is used both for instrumenting the solution program (fuzz driver) and mutator generation.
Each instrumented solution program is fed to \afl and runs for one hour.

We implement a prototype of \tool and provide 
    the scripts to run and collect experiment data, 
    publicly: \website.
We use Python as our main development language 
    and rely on the \texttt{perf} and \texttt{gcov}
    Linux utilities to collect instruction count, physical, 
    and code coverage metrics.
Overall, our artifact is implemented in approximately 10,000 LoC.

\subsection{Discussion on Measurements}
\label{appx:measurements}

To measure performance, we rely on the number of (retired) instructions based on physical execution time.
We use the \texttt{perf} Linux utility~\cite{weaver2013linux} 
    for both measurements.
While physical running time is a more intuitive measurement, 
    it can be prone to 
    interference from transient system effects—such as background 
    processes, scheduling policies, 
    and variable I/O latencies—which may mask true computational 
    cost~\cite{de2010new,zhang-eurosys13}.  
In contrast, measuring instruction counts is a metric that is more stable across run, platform-agnostic, 
  and is well understood to significantly correlate with executions 
  exhibiting performance bottlenecks~\cite{curtsinger-sosp15,zhang-eurosys13,iorgulescu-atc18,delimitrou-asplos14}.

Recently, LLM-based code analysis tools started to rely on instruction count measured through 
  hardware counters~\cite{liu-colm24, peng2025coffecodeefficiencybenchmark} or emulation~\cite{shypula-iclr24} 
  to evaluate code efficiency as it provides a more 
  reliable, low-variance 
  measurement than physical time alone.
Moreover, experiments in~\cite{peng2025coffecodeefficiencybenchmark} 
    found 
    instruction count measurements are approximately 
    1000$\times$ magnitude more stable than 
    physical execution time. 
This mirrors our own findings: our experiments reveal that
    physical time is $\sim$400$\times$ more variable than
    CPU instruction counts ($12\%$ versus $0.03\%$, 
    on average, see~\cref{sec:quantitative_results}
    and~\cref{appx:physycal-running-time}, respectively).

Prior works respectively rely on CPU simulators like Gem5~\cite{binkert2011gem5} (PIE~\cite{shypula-iclr24}), physical execution time (EffiBench~\cite{huang-neurips24-effibench}, Mercury~\cite{du2024mercury}), and hardware performance counters~\cite{wiki:Hardware_performance_counter} (EvalPerf~\cite{liu-colm24}, COFFE~\cite{peng2025coffecodeefficiencybenchmark}).
Gem5 is known to be stable, but the overhead is significant, and the CPU simulator does not necessarily reflect the physical performance.
% Physical execution time is most susceptible to the noise from environment, and the results should be interpreted with caution.
While physical running time offers an intuitive measure, it is susceptible to interference from transient system effects—such as background processes, scheduling variability, and I/O fluctuations—which can obscure true computational cost.
In contrast, hardware performance counters provide a more stable, low-noise, and platform-agnostic metric.
It records the number of executed instructions of program execution using the Linux \texttt{perf} tool~\cite{de2010new}.
It incurs low overhead and is highly reproducible.
Moreover, \citet{peng2025coffecodeefficiencybenchmark} demonstrates that it's linearly correlated with execution time with Pearson correlation coefficient of 0.96 \textasciitilde 1.0.
\section{Extended Results}

\subsection{Physical Running Time Measurements}
\label{appx:physycal-running-time}

\begin{table*}[!t]

\centering
\setlength{\tabcolsep}{16pt}
\renewcommand{\arraystretch}{1.1}
\small

\caption{\tool versus baselines (described in Section~\ref{sec:setup}) and its ablation.}

\begin{tabular}{lllr}
\toprule
\multirow{2}{*}{Technique} & \multicolumn{2}{c}{Execution time (ms)} & \multirow{2}{*}{Win rate}  \\
              & \multicolumn{1}{c}{Average} & \multicolumn{1}{c}{Median} &  \\
\midrule
\rowcolor[gray]{0.95}\multicolumn{4}{c}{\bf Comparison with baselines} \\

\tool & \textbf{228.60} & \textbf{82.33} & \textbf{60\%} \\
\tgprompt & 182.15 (\textcolor{red}{$\downarrow$}1.3$\times$) & 85.60 (\textcolor{green}{$\uparrow$}1.1$\times$) & 12\% \\
\perffuzz & 140.63 (\textcolor{red}{$\downarrow$}1.6$\times$) & 60.81 (\textcolor{red}{$\downarrow$}1.3$\times$) & 11\% \\
\evalperfslow & 139.83 (\textcolor{red}{$\downarrow$}1.6$\times$) & 61.69 (\textcolor{red}{$\downarrow$}1.3$\times$) & 8\%  \\
\evalperfrandom & 134.79 (\textcolor{red}{$\downarrow$}1.7$\times$) & 60.90 (\textcolor{red}{$\downarrow$}1.4$\times$) & 9\%  \\

\midrule

\rowcolor[gray]{0.95}\multicolumn{4}{c}{\bf Ablations} \\
\tool & \textbf{228.60} & \textbf{82.33} & \textbf{65\%} \\
\wedgenoinstrument & 171.63 (\textcolor{red}{$\downarrow$}1.3$\times$) & 71.54 (\textcolor{red}{$\downarrow$}1.2$\times$) & 29\% \\
\wedgedefaultmutator & \,\,\,86.57 (\textcolor{red}{$\downarrow$}2.6$\times$) & 59.13 (\textcolor{red}{$\downarrow$}1.4$\times$) & 4\% \\
\wedgeafl & \,\,\,78.27 (\textcolor{red}{$\downarrow$}2.9$\times$) & 45.13 (\textcolor{red}{$\downarrow$}1.8$\times$) & 2\% \\

\bottomrule
\end{tabular}

\label{tab:efficiency-running-time}

\end{table*}
% \FloatBarrier

We include physical running time measurements 
    as a reference. 
Note that the variance among the five repetitions is 
    more significant than for CPU instructions count, 
    namely $12\%$, on average.
While multiple factors contribute to this large variance,
    we argue that the dominant factor of ``noise'' is I/O 
    as a sizable fraction of programs 
    need to read thousands of KB, thus making them 
    I/O-bound.
Nevertheless, \cref{tab:efficiency-running-time} shows
    that our tool outperforms all other baselines 
    even when measuring physical execution time,
    a more noise-prone metric.

\subsection{Head-to-Head Numbers}
\label{appx:head-to-head}

Table~\ref{tab:histogram-raw-numbers} shows the 
    absolute numbers used to generate the histogram
    in~\cref{fig:histogram-head-to-head}.
Specifically, the values outside of parentheses 
    represent the number of programs that execute 
    more instructions when running \bench tests, 
    while those in parentheses indicate 
    the number of programs that execute more instructions 
    when running tests generated by one of the baselines 
    (each row).
Note that with one exception, \tgprompt tests that determine 
    programs to execute 10$\times$ more instructions, 
    our framework outperforms each technique.
Even so, \bench tests determine more programs for 
    longer by more than 50\% compared to \tgprompt 
    (\cref{tab:efficiency-main}).

\begin{table*}[!t]

    \centering
    \setlength{\tabcolsep}{4.5pt}
    \renewcommand{\arraystretch}{1}
    \small

  \caption{Source numbers for generating~\cref{fig:histogram-head-to-head}. Due to space constraints, we group numbers in bins of size 200\% (instead of 10\%) 
    into 6 larger buckets.
A bucket of \texttt{[x, y]} represents the number of programs 
    one technique incurs between x\% and y\% (inclusive) larger
    instruction count than the other.}
  \label{tab:histogram-raw-numbers}
  
\begin{tabular}{lrrrrrr}
\toprule
Techniques $\backslash$ Bins & [0, 199] & [200, 399] & [400, 599] & [600, 799] & [800, 999] & [1000,) \\
\midrule
\tool vs. & & & & & \\
\;\; \tgprompt & 17,117 ($\textcolor{red}{\downarrow}1,704$) & 1,524 ($\textcolor{red}{\downarrow}63$) & 158 ($\textcolor{red}{\downarrow}60$) & 148 ($\textcolor{red}{\downarrow}40$) & 155 ($\textcolor{red}{\downarrow}27$) & 311 ($\textcolor{green}{\uparrow}768$) \\
\;\; \perffuzz & 14,192 ($\textcolor{red}{\downarrow}1,758$) & 444 ($\textcolor{red}{\downarrow}83$) & 206 ($\textcolor{red}{\downarrow}57$) & 159 ($\textcolor{red}{\downarrow}\;\;3$) & 71 ($\textcolor{red}{\downarrow}\;\;0$) & 3,439 ($\textcolor{red}{\downarrow}\;\;17$) \\
\;\; \evalperfslow & 15,628 ($\textcolor{red}{\downarrow}1,856$) & 1,365 ($\textcolor{red}{\downarrow}69$) & 425 ($\textcolor{red}{\downarrow}43$) & 587 ($\textcolor{red}{\downarrow}\;\;7$) & 311 ($\textcolor{red}{\downarrow}11$) & 1,288 ($\textcolor{red}{\downarrow}454$) \\
\;\; \evalperfrandom & 15,879 ($\textcolor{red}{\downarrow}1,999$) & 966 ($\textcolor{red}{\downarrow}78$) & 588 ($\textcolor{red}{\downarrow}13$) & 664 ($\textcolor{red}{\downarrow}22$) & 367 ($\textcolor{red}{\downarrow}14$) & 1,004 ($\textcolor{red}{\downarrow}450$) \\
\bottomrule
\end{tabular}
\end{table*}
\FloatBarrier

\subsection{Effect of Input Size}
\label{appx:input-size-effect}

\begin{table*}[!t]

\centering
\setlength{\tabcolsep}{10pt}
\renewcommand{\arraystretch}{1}
\small

\caption{\tool versus length-stressing baselines (Section~\ref{sec:setup}) for problems whose inputs are below a threshold.}

\begin{tabular}{lllrrr}
\toprule
\multirow{2}{*}{Technique} & \multicolumn{2}{c}{\# of instructions ($\times10^8$)} & \multirow{2}{*}{Win rate} & \multicolumn{2}{c}{Slowdown over CC} \\
              & \multicolumn{1}{c}{Average} & \multicolumn{1}{c}{Median} & & \multicolumn{1}{r}{Average} & \multicolumn{1}{r}{Median} \\
\midrule
\rowcolor[gray]{0.95}\multicolumn{6}{c}{\bf Less than 1MB} \\

\tool & \textbf{4.29} & \textbf{0.17} & \textbf{59\%} & \textbf{113$\times$} & \textbf{1.11$\times$} \\
\tgprompt & 2.28 (\textcolor{red}{$\downarrow$}1.9$\times$) & 0.14 (\textcolor{red}{$\downarrow$}1.2$\times$) & 11\% & 75$\times$ & 1.01$\times$ \\
\evalperfslow & 2.57 (\textcolor{red}{$\downarrow$}1.7$\times$) & 0.07 (\textcolor{red}{$\downarrow$}2.3$\times$) & 14\% & 57$\times$ & 1.05$\times$ \\
\evalperfrandom & 2.59 (\textcolor{red}{$\downarrow$}1.7$\times$) & 0.08 (\textcolor{red}{$\downarrow$}2.2$\times$) & 16\% & 88$\times$ & 1.05$\times$ \\

\midrule
\rowcolor[gray]{0.95}\multicolumn{6}{c}{\bf Less than 100KB} \\

\tool & \textbf{3.71} & \textbf{0.08} & \textbf{56\%} & \textbf{88$\times$} & \textbf{1.06$\times$} \\
\tgprompt & 1.85 (\textcolor{red}{$\downarrow$}2.0$\times$) & 0.06 (\textcolor{red}{$\downarrow$}1.3$\times$) & 10\% & 60$\times$ & 1.00$\times$ \\
\evalperfslow & 2.48 (\textcolor{red}{$\downarrow$}1.5$\times$) & 0.03 (\textcolor{red}{$\downarrow$}2.3$\times$) & 16\% & 49$\times$ & 1.02$\times$ \\
\evalperfrandom & 2.52 (\textcolor{red}{$\downarrow$}1.5$\times$) & 0.03 (\textcolor{red}{$\downarrow$}2.5$\times$) & 18\% & 84$\times$ & 1.02$\times$ \\

\midrule
\rowcolor[gray]{0.95}\multicolumn{6}{c}{\bf Less than 10KB} \\

\tool & \textbf{3.10} & \textbf{0.02} & \textbf{55\%} & \textbf{20$\times$} & \textbf{1.02$\times$} \\
\tgprompt & 1.26 (\textcolor{red}{$\downarrow$}2.5$\times$) & 0.01 (\textcolor{red}{$\downarrow$}1.5$\times$) & 5\% & 9$\times$ & 1.00$\times$ \\
\evalperfslow & 1.79 (\textcolor{red}{$\downarrow$}1.7$\times$) & 0.01 (\textcolor{red}{$\downarrow$}1.4$\times$) & 19\% & 7$\times$ & 1.00$\times$ \\
\evalperfrandom & 1.53 (\textcolor{red}{$\downarrow$}2.0$\times$) & 0.01 (\textcolor{red}{$\downarrow$}1.4$\times$) & 21\% & 8$\times$ & 1.00$\times$ \\

\midrule
\rowcolor[gray]{0.95}\multicolumn{6}{c}{\bf Less than 1KB} \\

\tool & \textbf{2.98} & \textbf{0.01} & \textbf{54\%} & \textbf{5$\times$} & \textbf{1.01$\times$} \\
\tgprompt & 1.00 (\textcolor{red}{$\downarrow$}3.0$\times$) & 0.01 (\textcolor{red}{$\downarrow$}1.5$\times$) & 6\% & 1$\times$ & 1.00$\times$ \\
\evalperfslow & 1.63 (\textcolor{red}{$\downarrow$}1.8$\times$) & 0.01 (\textcolor{red}{$\downarrow$}1.3$\times$) & 16\% & 1$\times$ & 1.00$\times$ \\
\evalperfrandom & 1.45 (\textcolor{red}{$\downarrow$}2.1$\times$) & 0.01 (\textcolor{red}{$\downarrow$}1.3$\times$) & 24\% & 1$\times$ & 1.00$\times$ \\

\bottomrule
\end{tabular}

\label{tab:efficiency-input-size-variation}

\end{table*}
% \FloatBarrier

On the one hand, we compare our framework 
    with length-stressing baselines that are geared
    towards maximizing the input sizes~\cite{liu-colm24,peng2025coffecodeefficiencybenchmark}.
On the other, designing fuzzing tools that generate 
    performance-stressing tests are fundamentally 
    challenging, and requires specialized
    metrics and tailored mutators~\cite{lemieux_issta2018}.
Therefore, 
    we investigate if and how input size plays a role in the 
    quality of our tests (i.e., how ``performance-stressful'' 
    they are).

Specifically, we partition our original 
    dataset (Section~\ref{sec:setup}) into subsets of problems
    based on the maximum input size they require, and 
    compare \bench against tests generated by each baseline.
Table~\ref{tab:efficiency-input-size-variation} shows that,
    on average, our \tool outperforms the baselines when restricting input size.
The largest shift happens when compared with \tgprompt.
For problems whose inputs are less than 1KB, the difference 
    is almost double than comparing the two on the entire 
    dataset, 3.0$\times$ vs 1.5$\times$, respectively.
    
% While we are still inspecting individual cases, our preliminary investigation
%    suggests that this difference comes from the 
%     \tgprompt's focus on generating larger inputs with 
%     a bias towards universally edge-case patterns 
%     (e.g., near-complete graphs, sorted arrays).

%%% Bogdan: @Jun, I commented this part out as we might not 
%%% have time to add numbers here

% \subsection{Evaluating Effectiveness of Iterative Validator Generation}
% 

% \para{Accurary and Recall Analysis} We randomly sampled \placeholder{x} problems where a good validator is generated. We randomly select \placeholder{x} public/private inputs provided by \codecontest as positive samples and construct \placeholder{x} invalid inputs manually per problem. Then we validate the inputs with the validators. As shown in table \placeholder{x}, our Iterative Validator Generation achieves the precision and recall of \placeholder{x}, indicating it's a cheap and reliable alternative to manual annotation.

% \para{Effect of Iterative Refinement} To reveal the effect of iterative prompting we sampled \placeholder problems from \placeholder problems where a good validator was generated after multiple attempts, and Fig \placeholder shows the trend of recall \wrt number of iterations.

\subsection{Cost Analysis}

\begin{table}[!t]

\centering
\setlength{\tabcolsep}{10pt}
\renewcommand{\arraystretch}{1.1}
\small

\caption{Cost analysis of phases of \tool. }
\label{tab:token_cost_table}

\begin{tabular}{llll}
\hline
\textbf{Phase}                                              & \textbf{Tokens} & \textbf{Time (s)}           & \textbf{\# Executions} \\ \hline
Contrastive input pair mining                             & N/A             & 15.46                  & 99.65                  \\
Profiling feedback collection                             & N/A             & < 1                     & 2                      \\
PC constraint reasoning and constraint checker generation & 9633.16         & < 1                     & N/A                    \\
Constraint-aware mutator generation                       & 6905.27         & 180    & N/A                    \\
Constraint-guided fuzzing                                 & N/A             & 3600     & N/A                    \\ \hline
\end{tabular}
\end{table}

Table~\ref{tab:token_cost_table} presents the statistics of time or token cost in each phase.
The most time-consuming phase is constraint-guided fuzzing, which runs for one hour.
The most token-consuming phase is PC constraint reasoning and constraint checker generation, which consumes 9.6K tokens.
However, since we have 99.65 test inputs generated for each program on average, the time and token consumption per test input is low, \ie within 39 seconds and 166 tokens per input on average.

\section{Case Studies}
\label{appx:case-studies}

In this section, we consider several interesting case studies of 
    \pcc generated by \tool.
We first rank the synthesized constraints by the combined 
    size of their plain language description and corresponding
    generated checker code.
We then select five out of the top ten 
    problems that belong to different algorithmic classes.
We provide the full response produced by the \llm along with 
    additional selected case studies at~\website.

\subsection{Qualitative Analysis of Interesting Cases}
\label{appx:motivating-example}

In this section, we investigate the test generators synthesized
    by each baseline technique and compare them with the 
    \pcc generated by \tool, for Problem 633A which serves as 
    our motivating example.

\begin{lstinputlisting}
[language=Python, numbers=left, style=fancy, caption={\evalperfslow synthesized generator for Problem 633A}, label={lst:p633a-evalperfslow}, escapechar=!<>, lineskip=-1pt]
{listings/633_A_evalperf_slow_gen.py}
\end{lstinputlisting}

Inspecting the test generator synthesized by \evalperfslow 
    (Listing~\ref{lst:p633a-evalperfslow}) shows that it only generates inputs close to the problem's upper bounds (i.e., 99, 100, 10000) without even 
    utilizing the \codeIn{scale} parameter.

\begin{lstinputlisting}
[language=Python, style=fancy, numbers=left, style=fancy, caption={\evalperfrandom synthesized generator for Problem 633A}, label={lst:p633a-evalperfrandom}, escapechar=!<>, lineskip=-1pt, xleftmargin=1.5em, framexleftmargin=1.5em]
{listings/633_A_evalperf_random_gen.py}
\end{lstinputlisting}

The test generator synthesized \evalperfrandom (Listing~\ref{lst:p633a-evalperfrandom})
    random \codeIn{a} and \codeIn{b} within [1, 100] range 
    and a large \codeIn{c} (determined by \codeIn{scale}) 
    as the input.
Still, the effect is generating values close to the upper
    bounds of the problem.

% \begin{wrapfigure}{r}{0.4\textwidth}
\begin{lstinputlisting}
[language=Python, numbers=left, style=fancy, caption={\tgprompt synthesized generator for Problem 633A}, label={lst:p633a-tgprompt}, escapechar=!<>, lineskip=-1pt]
{listings/633_A_direct_prompting_gen.py}
\end{lstinputlisting}
% \end{wrapfigure}

\tgprompt (Listing~\ref{lst:p633a-tgprompt})   
    composes a series of concrete 
    test cases rather than synthesizing a 
    pattern-based generator.
While it attempts to implicitly reason about how to 
    synthesize performance-stressing inputs by including diverse patterns, some appear as generic corner cases.  
Thus, there is no guarantee these can trigger performance 
    bottlenecks.

In contrast, \tool relies on fuzzing with constraint-aware  
    mutators to efficiently search for a diverse set of 
    constraint-satisfying inputs.
Listing~\ref{lst:p633a-mutator} shows a constraint-aware mutator 
    for Problem 633A.
The mutator will mutate a previous input (seed input) to produce 
    the next input (mutated input).
The while loop will search the input space of (\codeIn{a, b, c}) 
    and ensure they conform to the constraints \codeIn{not abs (a - b) > 5 or c \% a == 0 or c \% b == 0}, \ie \codeIn{abs (a - b) <= 5 and c \% a != 0 and c \% b != 0} (already satisfying the first two out of three constraints in Listing~\ref{lst:p633a-constraint-checker}).

\begin{lstinputlisting}
[language=Python, style=fancy, numbers=left, style=fancy, caption={\tool synthesized constraint-aware fuzzing mutator for Problem 633A}, label={lst:p633a-mutator}, escapechar=!<>, lineskip=-1pt, xleftmargin=1.5em, framexleftmargin=1.5em]
{listings/633_A_solutions_0622_wedge_mutator.py}
\end{lstinputlisting}

\begin{lstinputlisting}
[language=C++, style=fancy, numbers=left, style=fancy, caption={\tool generated \pcc checker code}, label={lst:p633a-constraint-checker}, escapechar=!<>, lineskip=-1pt, xleftmargin=1.5em, framexleftmargin=1.5em]
{listings/633_A_solutions_0622_checker_2.cpp}
\end{lstinputlisting}
% \vspace{-1em}

\subsection{Other Representative Case Studies}

\para{\# Case 1: Problem 1209B, solution \#328.} Given a set of $n$ lights, each initially on/off and described by two parameters ($a_i$,$b_i$), light $i$ toggles its state at times $b_i$, $b_i$+$a_i$, $b_i$+$2a_i$, ... The goal is to find the maximum number of lights simultaneously on at any moment.

\begin{lstlisting}[language=C++, style=fancy, caption={Example from our dataset: problem 1209B, solution 328}, label={lst:p1209b-code}, 
escapechar=!]
#include <bits/stdc++.h>
using namespace std;
const long double pie = 3.14159265358979;
const long long mod = 1e9 + 7;
string vow = "aeiou";
void solve(int test_case) {
  int n;
  cin >> n;
  string s;
  cin >> s;
  int N = 1e3;
  vector<vector<int> > v(n, vector<int>(N, 0));
  vector<pair<int, int> > p(n);
  for (int i = 0; i < n; i++) cin >> p[i].first >> p[i].second;
  for (int i = 0; i < n; i++) {
    if (s[i] - '0') v[i][0] = 1;
    for (int j = 1; j < p[i].second; j++) v[i][j] = v[i][j - 1];
    int temp = v[i][0] ^ 1;
    for (int j = p[i].second; j < N; j += p[i].first) {
      for (int k = 0; j + k < N; k++) v[i][j + k] = temp;
      temp ^= 1;
    }
  }
  int ans = 0;
  for (int j = 0; j < N; j++) {
    int temp = 0;
    for (int i = 0; i < n; i++) temp += v[i][j];
    ans = max(ans, temp);
  }
  cout << ans;
  cout << "\n";
}
int main() {
  ios_base::sync_with_stdio(false);
  cin.tie(0);
  cout.tie(0);
  int t = 1;
  for (int i = 0; i < t; i++) solve(i);
  return 0;
}
\end{lstlisting}

The code in Listing~\ref{lst:p1209b-code} simulates a fixed
    time window $N=1,000$, instead of reasoning 
    analytically about the periodic patterns which 
    requires solving a least‐common‐multiple type of 
    problem.
Specifically, the program builds a two-dimensional 
    array where for each light $i$ it first copies its 
    initial state up to $t=b_i$, then for each toggle 
    epoch $t=b_i$,$b_i+a_i$,... it fills the remainder 
    of the row in one shot via a nested ``for $k$'' 
    loop at line $20$. 
Finally, it scans each column $t$, sums up 
    $v[i][t]$ for $i=1,n$, and tracks the maximum (line $28$).
In the worst case, when $a_i=1$, the program performs 
    $O(N^2)$ iterations per toggling operation, for an overall complexity of $O(n\cdot N^2+n\cdot N)$.

\begin{lstlisting}[language=C++, style=fancy, caption={PC-constraints as C++ checker functions: problem 1209B, solution 328}, label={lst:p1209b-constraints}, 
escapechar=!]
void check_small_a_values(const vector<pair<int, int>>& p) {
    int small_a_count = 0;
    for (const auto& pair : p) {
        if (pair.first <= 2) {  // Assume 'small' a_i values are <= 2
            small_a_count++;
        }
    }
    if (small_a_count > 50) {  // Arbitrary threshold, adjust as needed
        cerr << "Warning: Performance bottleneck condition triggered - many lights have small 'a' values" << endl;
    }
}

void check_synchronized_b_values(const vector<pair<int, int>>& p) {
    map<int, int> b_count;
    for (const auto& pair : p) {
        b_count[pair.second]++;
    }
    for (const auto& [b_value, count] : b_count) {
        if (count > 30) {  // Arbitrary threshold for synchronization
            cerr << "Warning: Performance bottleneck condition triggered - synchronized 'b' values" << endl;
        }
    }
}

void check_large_number_of_lights(int n) {
    if (n > 90) {  // Close to the upper constraint
        cerr << "Warning: Performance bottleneck condition triggered - high number of lights" << endl;
    }
}
\end{lstlisting}

\tool identifies three types of pc-constraints for this program, listed in \ref{lst:p1209b-constraints}.

\begin{enumerate}
\item The first performance-characterizing constraint (i.e.,
    \texttt{check\_small\_a\_values}) finds that if 
    many lights have very small $a_i$ (e.g. 1), 
    translates to more iterations of the outer loop
    at line $15$.
    Consequently, each iteration invokes a full 
    inner copy (loop at line $20$). 
    Thus, small periods significantly amplify the work performed 
    by the nested‑loop work.
\item The second performance-characterizing constraint (i.e.,
    \texttt{check\_synchronized\_b\_values}) points to 
    clusters of $b_i$ values: 
    if many lights share the same or 
    close $b_i$, their current state lineup, causing 
    the code to execute the heavy inner loop at line 
    $20$ for multiple lights at the same early offsets.
    Specifically, low or repeated \texttt{p[i].second} forces expensive fill operations to execute immediately and for many lights in fast succession.
\item Finally, the third performance-characterizing constraint (i.e.,
    \texttt{check\_large\_number\_of\_lights})
    simply indicates that as $n$ approaches 
    its upper bound near $100$, 
    the total nested work scales linearly in $n$. 
    Each additional light multiplies the cost of the 
    $O(N^2)$ toggling loops and the $O(N)$ 
    scan across time.
\end{enumerate}

A purely specification or problem-statement based
    performance analysis might determine that 
    small toggle periods and a large number of lights
    trigger multiple loop iterations because
    any simulation of a periodic, large number of 
    events would exhibit that. 
However, the actual performance profile of the 
    code actual performance relies on two highly 
    implementation‐specific choices.
First, instead of toggling cell by cell, the 
    implementation writes an entire suffix of 
    the time‐array (for loop at line $20$), thus
    a light toggle to linear ($O(N)$) instead 
    of constant operation.
Second, the choice of simulating the maximum number of 
    seconds possible, $1,000$ $1,000$, irrespective of the 
    input.
An optimal solution would take into account that each 
    light's behavior is periodic, namely once it 
    reaches its first toggle at $t=b_i$, thereafter it 
    repeats every $a_i$ seconds.
This, naturally, translates into cycles of length equal
    to the \textit{lowest common multiplier}:
    $\ell = \texttt{lcm}(a_1,a_2,\ldots,a_n)$.
The complexity of the optimal program is, therefore,
    approximately two orders of magnitude since,
    based on the problem specifications, 
    $1\leq a_i, b_i \leq 5$. 
Since $\ell \leq lcm(1, 2, 3, 4, 5) = 60$ this 
    leads to an optimal complexity of 
    $O(\ell^2 \cdot n)=O(3,600\cdot n)$ instead of the 
    current $O(N^2 \cdot n)=O(10^6 \cdot n)$.

\para{\# Case 2: Problem 1118D1, solution \#30.} Given a homework of $m$ pages 
    and $n$ coffee cups with caffeine doses $a_1$, ..., $a_n$, 
    the problem asks to compute the minimum number of days a 
    student must schedule to drink those cups so the pages 
    he writes reach or exceed m.
When the student drinks $k$ cups on a single day labeled 
    $a_{i_1}$, ..., $a_{i_k}$, the first cup lets him write $a_{i_1}$ pages, the second max(0, $a_{i_2}$ – 1), 
    the third max(0, $a_{i_3}$ – 1), and so on. 
Thus, the task is to find the smallest such sequence given 
    $n$, $m$, and the list of caffeine values.

\begin{lstlisting}[language=C++, style=fancy, caption={Example from our dataset: problem 1118D1, solution 30}, label={lst:p1118d1-code}, 
escapechar=!]
#include <bits/stdc++.h>
using namespace std;
const long long int N = 1e6;
long long int n, m;
vector<long long int> v(N);
bool check(long long int days) {
  long long int pages = 0, k = 0;
  for (long long int i = 0, j = 0; j < n; i++, j++) {
    pages += max(0ll, v[j] - k);
    if (i + 1 == days) i = -1, k++;
  }
  return pages >= m;
}
int32_t main() {
  scanf("%lld", &n);
  scanf("%lld", &m);
  long long int sum = 0;
  for (long long int i = 0; i < n; ++i) {
    scanf("%lld", &v[i]);
    sum += v[i];
  }
  if (sum < m) {
    printf("-1");
    return 0;
  }
  sort(v.begin(), v.end(), greater<long long int>());
  long long int ans = 1e16;
  long long int low = 1;
  long long int high = n;
  while (low <= high) {
    long long int mid = (low + high) / 2;
    if (check(mid)) {
      ans = min(ans, mid);
      high = mid - 1;
    } else {
      low = mid + 1;
    }
  }
  printf("%lld", ans);
  return 0;
}
\end{lstlisting}

The program in Listing~\ref{lst:p1118d1-code} implements a binary 
    search on the expected output, $m$.
For each value of $m$, the binary search simulates writing pages over 
    $days$ days (function \texttt{check}) by iterating 
    each cup exactly once, decrementing future cups by an increasing 
    offset $k$ whenever a day's quota is reached at line  
    (lines $9$-$10$).

When the binary search checks a candidate value, 
    it always scans all $n$ cups in the for loop at line 8, 
    performing $O(n)$ work per invocation.
The binary search interval shrinks slowly, 
    forcing multiple calls to check, which runs in $O(n)$ time. 
Thus, when $n$ approaches its upper limit and $m$ 
    nearly equals the sum of all $a_i$, the combination of an 
    $O(\log n)$ binary search multiplied by an linear check 
    becomes the dominant bottleneck.

\begin{lstlisting}[language=C++, style=fancy, caption={Performance-characterizing constraints as C++ checker functions: problem 1118D1, solution 30}, label={lst:p1118d1-constraints}, 
escapechar=!]
void check_binary_search_invariant(long long sum, long long m, int search_iterations) {
    if (search_iterations > 100 && sum >= m && (m > 0.9 * sum)) {
        cerr << "Warning: binary_search_invariant triggered - extended binary search due to close capacity and requirement" << endl;
        abort();
    }
}

void check_cup_order_invariant(const vector<long long>& v, long long m) {
    long long potential_pages = 0;
    int decrement_operations = 0;
    for (size_t i = 0; i < v.size(); ++i) {
        potential_pages += max(0ll, v[i] - (long long)i);
        if (v[i] > (long long)i) {
            ++decrement_operations;
        }
    }
    if (decrement_operations > 50 && potential_pages < m) {
        cerr << "Warning: cup_order_invariant triggered - extensive decrement operations" << endl;
    }
}
\end{lstlisting}

\tool identifies two types of \pcc for this program, listed in Listing~\ref{lst:p1118d1-constraints}.

\begin{enumerate}
\item The first performance-characterizing constraint
    (\texttt{check\_binary\_search\_invariant}) finds that the binary search becomes expensive by performing a large number of iterations 
    when the necessary pages $m$ are close to the cumulative caffeine 
    sum of all cups $n$ and $n$ is particularly large.
\item The second performance-characterizing constraint 
    (\texttt{check\_cup\_order\_invariant}) finds that the
    loop in \texttt{check} (lines $8$-$10$) performs on the order 
    of $n \times days$ operations. 
This happens because \texttt{check} iterates over the entire list of 
    $n$ cups adding $\texttt{max}(0, v[j]-k)$ per page and 
    resetting the loop counter $i$ every $days$ 
    iterations. 
\end{enumerate}

The first performance-characterizing constraint relates to binary search complexity and is 
    generic since searching over a large range is typically worst-case 
    in terms of time complexity, the target value lies in the middle, 
    forcing the search to ``zig‑zag'' and perform a large number of 
    iterations. 

In contrast, the second performance-characterizing constraint is implementation-specific. The LLM reasons about how the check loop resets $i$ and increments $k$ to model diminishing returns and then calculates again 
    $max(0, v[j]-k)$ for every cup.
Note that the LLM borrows the same loop structure as in the
    original \texttt{check} function (e.g. reset of $i$ and 
    increment of $k$) so the checker code faithfully reproduces the 
    slowdown pattern.

\para{\# Case 3: Problem 546C, solution \#567}. Given a deck of $n$ distinct 
    cards split arbitrarily into two decks, one per player), the 
    problem asks to simulate a game in which, in each round, both 
    players draw their top card, and the player with the higher value 
    takes the opponent's card first and then their own, placing both at the bottom of their stack. 
When one player's stack becomes empty, the other wins.

\begin{lstlisting}[language=C++, style=fancy, caption={Example from our dataset: problem 546C, solution 567}, label={lst:546c-code}, 
escapechar=!]
#include <bits/stdc++.h>
using namespace std;
queue<int> r1, r2;
int n, x, TLE, asd;
bool flag;
int main() {
  cin >> n >> x;
  for (int i = 1; i <= x; i++) cin >> asd, r1.push(asd);
  cin >> x;
  for (int i = 1; i <= x; i++) cin >> asd, r2.push(asd);
  while (TLE < 10000000) {
    if (r1.size() == 0 || r2.size() == 0) {
      flag = 1;
      break;
    }
    TLE++;
    int u = r1.front(), v = r2.front();
    r1.pop(), r2.pop();
    if (u > v)
      r1.push(v), r1.push(u);
    else
      r2.push(u), r2.push(v);
  }
  if (flag)
    if (r1.size() == 0)
      cout << TLE << " " << 2 << endl;
    else
      cout << TLE << " " << 1 << endl;
  else
    puts("-1");
  return 0;
}
\end{lstlisting}

The program in Listing ~\ref{lst:546c-code} simulates the game by
    representing the player as two queues.
In each round, the implementation pops the head of each queue, 
    compares the two values and adds both elements to the 
    queue with the highest value of the two.
If either queue is empty, the program terminates.
While straightforward, the implementation has two significant flaws.
First, the code does not store past game states after each round.
This leads the simulation to repeat the same deck configurations 
    repeatedly until the $TLE=10^6$ threshold is met instead of 
    breaking early if a cycle is detected, thus wasting many CPU cycles.
Second, if the largest cards in either deck differ by a small amount 
    (e.g., $\leq 1$) the main loop will iterate for significantly 
    more trials.
When neither player's top card is consistently stronger, the 
    \texttt{pop}/\texttt{push} operations keep both queues nearly 
    the same size. 
This balanced exchange prevents either queue from becoming empty
    quick, forcing multiple extra iterations of the same
    \texttt{pop}/\texttt{push} logic.

\begin{lstlisting}[language=C++, style=fancy, caption={PC-constraints as C++ checker functions: problem 546C, solution 567}, label={lst:546c-constraints}, 
escapechar=!]
// Check if the queue configurations have repeated, indicating a cycle
void check_cycle_invariant(const queue<int>& r1, const queue<int>& r2, unordered_set<string>& seenConfigurations) {
    string config;
    auto copy1 = r1, copy2 = r2;
    while (!copy1.empty()) {
        config += tostring(copy1.front()) + ",";
        copy1.pop();
    }
    config += "|";
    while (!copy2.empty()) {
        config += to_string(copy2.front()) + ",";
        copy2.pop();
    }
    
    if (seenConfigurations.find(config) <> seenConfigurations.end()) {
        cerr << "Warning: Cycle detected - potential infinite loop" << endl;
    }
    seenConfigurations.insert(config);
}

// Check if neither player can win quickly due to balanced card strength
void check_balance_invariant(const queue<int>& r1, const queue<int>& r2) {
    int max1 = *max_element(r1.front(), r1.back());
    int max2 = *max_element(r2.front(), r2.back());
    if (abs(max1 - max2) <= 1) { // Arbitrarily chosen threshold for similar max strength
        cerr << "Warning: Balanced card strength - prolonged game possible" << endl;
    }
}

// Check for excessive number of rounds
void check_excessive_rounds(int TLE) {
    if (TLE > 1000) { // Example threshold, can be adjusted for practical purposes
        cerr << "Warning: Excessive number of game rounds" << endl;
    }
}
\end{lstlisting}

\tool identifies three types of performance-characterizing constraints for this program, listed in Listing~\ref{lst:546c-constraints}.

\begin{enumerate}
\item The first performance-characterizing constraint 
    (\texttt{check\_cycle\_invariant}) checks whether the distribution
    of the two decks is prone to repeated states.
The LLM detects that the original implementation does not take into 
    account repetitions to terminate the main loop thus inputs with 
    pathology is likely to force the loop at line $11$ to iterate 
    until reaching the $TLE=10^6$ threshold wasting unnecessary cycles.
\item The second performance-characterizing constraint 
    (\texttt{check\_balance\_invariant}) checks for ``back‑and‑forth'' \texttt{push} operation which cause minimal net change 
    in queue size and prolong the game.
The inefficiency comes from the two \texttt{push} calls per round 
    (lines $20$ and $22$) and the fact that the losing card is enqueued first. 
This particular ordering choice leads to ``reversing'' of the first player's win since eventually, the smaller card, which was 
    enqueued first will move back to the second player, 
    canceling the gains of the first one. 
\item The third performance-characterizing constraint 
    (\texttt{check\_excessive\_rounds}) simply reflect a generic 
    and straightforward intuition: more loop iterations means 
    more CPU cycles and more running time.
\end{enumerate}

While the third performance-characterizing constraint is generic and could have been synthesized by an LLM without reasoning about the code, the first two are implementation-specific.
Any queue-based simulation with a bounded state that can revisit prior configurations is prone to cycles.
However, this is problematic only when such a program does not track repeated states, which is precisely what happens in this case.
Moreover, the ``back-and-forth'' card exchange property is highly specific to the program the LLM is reasoning about.
It happens precisely because of the choice of enqueuing order.
Should the implementation enqueue the two values in the opposite order, it would be unlikely that there is any observable ``back-and-forth'' where cards move from one queue to another and then back again.

\para{\# Case 4: Problem 16B, solution \#34.} Given a set of 
     $m$ containers, where the i-th container holds $a_i$ match boxes, each containing $b_i$ matches, 
     the goal is to select up to $n$ boxes (without splitting boxes) to carry in a backpack so as to maximize the total number of matches carried away.

\begin{lstlisting}[language=C++, style=fancy, caption={Example from our dataset: problem 16B, solution 34}, label={lst:p16b-code}, 
escapechar=!]
#include <bits/stdc++.h>
using namespace std;
long long sumofdigits(string s) {
  long long sum = 0;
  for (long long i = 0; i < s.size(); i++) {
    int digit = s[i] - '0';
    sum += digit;
  }
  return sum;
}
int main() {
  int n;
  vector<pair<int, int>> v;
  cin >> n;
  int m;
  cin >> m;
  for (int i = 0; i < m; i++) {
    int x, y;
    cin >> x >> y;
    pair<int, int> p(x, y);
    v.push_back(p);
  }
  int sum = 0;
  for (int i = 0; i < v.size() - 1; i++) {
    for (int j = i + 1; j < v.size(); j++) {
      if (v[j].second > v[i].second) {
        pair<int, int> p = v[i];
        v[i] = v[j];
        v[j] = p;
      }
    }
  }
  int ans = 0;
  for (int i = 0; i < v.size(); i++) {
    int counter = 0;
    if (sum == n) {
      break;
    }
    int t = n - sum;
    while (counter < v[i].first && t--) {
      counter++;
      sum++;
      ans += v[i].second;
    }
  }
  cout << ans << endl;
  return 0;
}
\end{lstlisting}

The code in Listing~\ref{lst:p16b-code} performs two stages to 
    solve the problem.
First, it performs a simple $O(m^2)$ sort (lines $24$-$32$) that 
    orders the containers by their matches per box $b_i$ 
    descendingly.
Second, it tries to fill the backpack in a greedy fashion 
    (lines $34$-$45$), as follows: 
For each container in sorted order, it enters a \texttt{while} 
    loop (line 40) picks boxes one by one, decrementing the 
    remaining capacity per box until either the 
    container’s supply is exhausted, or the capacity 
    aggregator variable $sum$ reaches $n$.
Note that before entering the inner loop 
    decrements $t$ exactly once per box being packed (line $39$).
Thus a large $n$ causes the code to run one 
    iteration for each available box across all containers. 
Moreover, if $n$ is much larger than the sum of all box 
    capacities, the code still visits every box which 
    yields a computational cost linear in the total capacity.
Also, the code does not appear to terminate early if the 
    box supply is exhausted.
The code breaks out of the outer loop only if the 
    current aggregated box capacity 
    is equals to $n$ (lines $36$-$37$).
If the supply is smaller, however, the outer loop iterates
    unnecessarily through all containers.

\begin{lstlisting}[language=C++, style=fancy, caption={PC-constraints as C++ checker functions: problem 16B, solution 34}, label={lst:p16b-constraints}, 
escapechar=!]
void check_large_n_invariant(int n, int m, const vector<pair<int, int>>& v) {
    long long totalBoxes = 0;
    for (const auto& container : v) {
        totalBoxes += container.first;
    }
    if (n > 10 * totalBoxes) {
        cerr << "Warning: Performance bottleneck condition triggered - n is much larger than available matchboxes" << endl;
    }
}

void check_small_total_boxes_invariant(int n, int m, const vector<pair<int, int>>& v) {
    long long totalBoxes = 0;
    for (const auto& container : v) {
        totalBoxes += container.first;
    }
    if (totalBoxes < n / 10) {
        cerr << "Warning: Performance bottleneck condition triggered - insufficient matchboxes compared to n" << endl;
    }
\end{lstlisting}

\tool identifies two types of pc-constraints for this program, listed in \ref{lst:p1209b-constraints}.

% \begin{enumerate}
% \item 
The first performance-characterizing constraint 
    (\texttt{check\_large\_n\_invariant}) detects 
    when the target capacity $n$ vastly exceeds the 
    total number of available boxes.
    The inner loop (lines $40$-$44$) executes exactly one iteration per box taken.
    So, the mode infers that when $n$ is significantly larger than the available boxes,
    the code performs $O(\sum_i a_i)$ iterations since each time
    it decrements $t$ by $1$.
% \item

The second performance-characterizing constraint (\texttt{check\_small\_total\_boxes\_invariant}) detects when the aggregate supply of boxes is substantially less than $n$, since the loop break condition (line $37$) applies only when the bag is exactly full.
Otherwise, the outer loop (line $34$) iterates over all containers, wasting cycles when capacity is no longer available.
% \end{enumerate}

A purely specification or problem-statement based
    performance analysis could infer that larger $n$ could 
    cause longer execution, and that a typical implementation 
    would iterate until the capacity is full or exceeded.
However, the first constraint hinges on the implementation 
    choice of simulating each box by decrementing $t$.
This is suboptimal because to solve the problem, the algorithm
    only needs to know how many boxes to take, not to process
    them individually.
Similarly, the second constraint speculates that the code does not exit immediately once it is determined that the matchbox supply is depleted before reaching $n$.

% \subsection{TODO....}
% Example product execution feedback (coverage and hit count) in Listing~\ref{lst:p633a-code}.
% \noindent
% % \begin{wrapfigure}{r}{0.6\textwidth}
% \begin{lstinputlisting}
% [language=C++, numbers=none, caption={Contrasting feedback collected from contrasting inputs on the solution program in Figure~\ref{fig:motivating}}, label={lst:p633a-prodoct-cov}, escapechar=!<>, lineskip=-1pt]
% {listings/633_A_solutions_0622.cov}
% \end{lstinputlisting}
% \end{wrapfigure}
\section{Broader Impact}
\label{sec:discussion}

% \paragraph{Limitations.}
% First, our approach incorporates prompting and fuzzing, thus incurring considerable token cost and execution time.
% Second, the performance-characterizing constraint reasoning depends on mining high-quality slow-fast input pairs, thus requiring either a large corpus of test inputs or an additional step to create a number of inputs.
% Third, the process of searching for constraint-satisfying inputs depends on the fuzzing engine, e.g., AFL++~\cite{fioraldi-woot20} can only mutate an input file whose size is smaller than a threshold, thus cannot handle large inputs.
% If a constraint is complex and deeply nested, the fuzzing engine might not be strong enough to satisfy the constraint within the given time budget.

% \paragraph{Broader Impact.}
The positive impact of our research includes the following.
% \begin{itemize}
    % \item 

First, we release a performance test benchmark that can evaluate the efficiency of LLM-generated code and performance-improving code edits, which can facilitate future research.

Second, we develop a methodology to generate stress tests by combining the advantage of the reasoning ability of LLMs and the searching ability of fuzzing, which can inspire future research in performance test generation on real-world software.

Third, instead of introducing more ambitious task formulations and calling for new LLM agentic workflows, we advocate constraining the role of LLMs in system reliability and security applications.
We hope the general paradigm described in this paper, i.e., generating code specifications to interact with existing expert-developed program reasoning tools, can help the community rethink how LLMs should participate in critical applications.
% \end{itemize}

The concrete negative impact is our approach could be potentially used by malicious attackers to curate stressing inputs to perform Denial-of-Service (DoS) attacks~\cite{petsios_sigsac17}. 
We aim to study the positive usage of our approach, e.g., finding performance issues in large-scale systems, to find and mitigate these potential vulnerabilities.
% However, as our approach requires a large amount of diverse test inputs (along with profiling information) and source code, which are usually not available, the risk of performing this attack is low.

\newpage
% \newpage
\section{Prompts}
\label{appx:prompts}
% \begin{wrapfigure}{r}{0.54\textwidth}
% \begin{minipage}{0.50\textwidth}
% \begin{lstinputlisting}
% [language=Python, numbers=left, caption={\evalperfrandom synthesized generator for Problem 633A}, label={lst:p633a-evalperfrandom}, escapechar=!<>, lineskip=-1pt]
% {listings/633_A_evalperf_random_gen.py}
% \end{lstinputlisting}
% \end{minipage}
% \end{wrapfigure}

\subsection{Performance-Characterizing Constraint Reasoning Prompt}

\noindent\fbox{%
\parbox{\textwidth}{%
\small
\textbf{(A) Context}

You are an experienced C software engineer focusing on performance bottlenecks. You have:\\
1. A problem statement describing a task or algorithm (with constraints such as $n \leq 100$).\\
2. A C program that implements a solution to that problem.\\
3. Two inputs: a ``fast'' input that completes quickly, and a ``slow'' input that takes much longer---both inputs have similar size/structure.\\
4. Line-level hit counts for both runs, showing which lines get hit significantly more often on the slow input.

Your goal is to diagnose why the program runs slowly for the slow input and derive conditions or invariants that capture what triggers this slowdown.

\mbox{}\\ 
\textbf{(B) Tasks:} Analyze the given code and generate performance-characterizing invariants in natural language

\textbf{Phase 1:} Identify expensive or inefficient code fragments.\\
1. Compare line-level hit counts between the fast and slow runs.\\
2. Pinpoint lines or functions that get significantly more hits under the slow input.\\
3. Infer how these lines might be interacting with data structures, loops, recursion, etc., especially as they relate to the input constraints (e.g., $n \leq 100$).

\textbf{Phase 2:} Derive performance-characterizing invariants (natural language).\\
1. Generate natural language statements that describe conditions under which the program likely enters a slow path.\\
2. Avoid using specific numeric values from the slow input; abstract them into categories or thresholds. However, make sure those thresholds adhere to the input constraints of the problem.\\
3. Correlate these conditions strongly to input patterns (e.g., ``when $n$ is close to 100 and there is a nested loop,'' or ``when a data structure is repeatedly sorted'').\\
4. Ensure your statements are broad enough to catch possible future slow scenarios, but still reflect realistic triggers given the constraints (like $n \leq 100$).

Note that not all performance-characterizing invariants are about maximising input size. You may refer to the following examples for inspiration --- some maximising the input size, some not --- but do not simply replicate them exactly. Rather, use them as inspiration to infer and tailor performance-characterizing invariants tailored for the C code and problem statement you were asked to analize:

% --- EXAMPLES OMITTED FOR BREVITY IN LATEX SNIPPET ---

(Include the same examples you have, with indentation or \\ to split lines appropriately.)

\mbox{}\\ 
\textbf{(C) Output Requirements}\\
1. Provide a list of natural language performance invariants explaining under what conditions the code slows down.\\
2. Do not mention or rely on exact values from the provided slow input.\\
3. Use or suggest threshold values that align with problem constraints (e.g., $n \leq 100$).\\
4. The output should be a concise, descriptive set of statements about performance triggers.

\mbox{}\\ 
\textbf{(D) Important Considerations}\\
1. Avoid hardcoding. Don’t rely solely on the exact values from the provided slow input; think in terms of categories or thresholds that lead to slow execution.\\
2. Avoid checks inside tight loops. Place checks in a way that does not significantly degrade performance.\\
3. Focus on fuzzer utility. The checks should help a fuzzer detect slow performance triggers by hitting these conditions.

\mbox{}\\ 
\textbf{(E) Problem Statement}\\
\placeholder{problem\_statement}

\vspace{0.5em}
\textbf{(F) Program Solving the Problem Statement}\\
\placeholder{one\_solution}

% \mbox{}\\ 
\vspace{0.5em}
\textbf{(G) The Slow \& Fast Inputs}\\
\textbf{(G.1) Slow Input}\\
\placeholder{slow\_input}\\
\textbf{(G.2) Fast Input}\\
\placeholder{fast\_input}

% \mbox{}\\ 
\vspace{0.5em}
\textbf{(H) Hit Count Information of Slow Input and Fast Input (Aggregated):}\\
\placeholder{product\_cov}

}
}

\newpage
\subsection{Constraint Checker Generation Prompt}
% \begin{figure}[t]
\noindent\fbox{%
\parbox{\textwidth}{%
\small
\textbf{(A) Context}

You have already:\\
1. Identified expensive code fragments (Phase 1).\\
2. Derived performance-characterizing invariants in natural language (Phase 2).

Now, you \textbf{MUST} transform these invariants into runtime checks and integrate them into the given C++ program.

\mbox{}\\ 
\textbf{(B) Tasks:} Revisit the performance-characteristic invariants you inferred in natural langauge and write faithful, error-free C++ code snippets to implement them.

You \textbf{MUST} perform this task in two phases and provide separate answers for both: First, translating the invariants into checker code in C++ (phase 3, below). Second, integrating those checker C++ code snippets with the original program for which you inferred those invariants (phase 4, below).   

\mbox{}\\ 
\textbf{Phase 3:} Implement the natural language invariants inferred previously, in C++. In this phase you are asked to,\\
1. For each natural language invariant from Phase 2, you \textbf{MUST} produce C++ code that checks the condition at runtime.\\
2. You \textbf{MUST NOT} relax or trivialize the checker code implementing these performance-characterizing invariants. You \textbf{MUST} faithfully implement them as described.\\
3. Use the following template for writing checker code in C++ which could also be implemented as a helper function:

\texttt{if (/* condition based on the NL invariant */) \{\\
\hspace*{1em}cerr << "Warning: Performance bottleneck condition triggered!" << endl;\\
\hspace*{1em}abort();\\
\}}

\mbox{}\\ 
Note that not all performance-characterizing invariants are about maximising input size. You may refer to the following examples for inspiration --- some maximising the input size, some not --- but do not simply replicate them exactly. Rather, use them as inspiration to infer and tailor performance-characterizing invariants tailored for the C++ code and problem statement you were asked to analize:

// in-context examples......

\mbox{}\\ 
\textbf{Phase 4:} Propagate and insert conditional checks. In this phase you are asked to,\\
1. Place each check at an effective point in the control/data flow (e.g., after reading inputs, before heavy loops) so you do not add overhead in tight loops. Note that the checker code could also be implemented as a helper function.\\
2. If multiple checks overlap, merge or adjust them carefully to avoid redundant warnings.\\
3. Provide the final, instrumented C++ code in code fences. Ensure it compiles cleanly and runs without errors.\\
4. For each inserted check, add a short comment explaining which bottleneck it detects.

\mbox{}\\ 
Note the following important considerations when translating the inferred performance-characterizing invariants into code and propagating the checkers to the most effective program point by instrumenting the original code:\\
1. Avoid hardcoding. Don’t rely solely on the exact values from the provided slow input; think in terms of categories or thresholds that lead to slow execution.\\
2. In addition to the warning message, remember to insert an \texttt{abort()} statement at the end of the checker.\\
3. Focus on fuzzer utility. The checks should help a fuzzer detect slow performance triggers by hitting these conditions.

\textbf{As a refresher:} below you are provided with the same program statement and C++ code for which you already inferred performance-characterizing invariants:

\texttt{Problem statement:} \placeholder{problem\_statement}\\
\texttt{Solution (C++ code):} \placeholder{solution}
}
}
% \caption{The prompt used for transforming natural language invariants into runtime checks.}
% \end{figure}
\newpage

\subsection{Performance-Constraint-Aware Mutator Generation Prompt}
\noindent\fbox{%
\parbox{\textwidth}{%
\small

\mbox{}\\
Here is an example custom mutator provided in AFL++ repo:

\placeholder{mutator\_example}

\mbox{}\\
Could you learn from it and generate a new one?

\mbox{}\\
The constraints of input are described here:

\textbf{-------- Problem begins}

\placeholder{problem\_statement}

\textbf{-------- Problem ends}

\mbox{}\\
Here are some example inputs that you can use as a reference:

\placeholder{reference\_inputs}

\mbox{}\\
Here are the performance related conditions summarized by another LLM, e.g., \texttt{check\_perf\_condition\_*(condition)}. These conditions are believed to be related to performance of the solution, i.e., when the conditions are satisfied, the solution will likely run slower than when they are not satisifed. Please learn from them so that the generated mutator can produce more inputs to satisfy the conditions so that the generated inputs can make the solution run slower.

\mbox{}\\
\textbf{-------- Constraints summary begins}

\placeholder{constraints\_content}

\textbf{-------- Constraints summary ends}

\mbox{}\\
As the context, here is the solution program (along with coverage and hit count information under slow/fast inputs) where the constraints are generated on:

\placeholder{product\_cov\_content}

\mbox{}\\
Please note that:
1. You should ensure the mutated inputs follow the input constraints as much as possible. Note that 100000(10\textasciicircum5) is usually written as 105, 1000000(10\textasciicircum6) is usually written as 106, etc.\\
2. You should implement an input generator and incorporate it in the mutator, so that each iteration it can randomly choose to mutate the last input or use the generator to generate inputs from scratch. Note that you should avoid using random generator with a large range of values when generating the size input (e.g., length of array, etc.), e.g., \texttt{random.randint(1, 10000)} (say 10000 is the upper bound), as it may not stress the program enough. Instead, use values that are more likely to cause inefficiencies, for example, values that are same or closer to the upper bound, like 10000, 9999, 9990 or \texttt{random.randint(9900, 10000)}, etc. Feel free to directly use the upper bound as the size of the input. But for numbers that are not the size of the input, you may want to generate random values to improve the input diversity and cover different patterns.\\
3. You should try to implement multiple mutation operations that can be randomly selected. You can implement mutation operations that could potentially explore corner cases of the program (e.g., upper/lower bound).\\
4. You can add a try catch block to the mutation module so that if the mutation failed you can fall back to the generator since the generator is believed to be more robust.\\
5. Please make sure to use \texttt{bytearray(str, 'utf-8')} to transform string to bytearray, instead of \texttt{str.encode()} as we need mutable objects; the latter will produce immutable objects.

}
}
\newpage
\subsection{Baseline \tgprompt Prompt}
\noindent\fbox{%
\parbox{\textwidth}{%
\small

\mbox{}\\
You are an experienced Python software engineer. Your task is to produce a test generator in the form of a Python program that, based on the input specifications in the problem statement, generates tests that exhaust the solutions of the problem more.

\mbox{}\\
Problem Statement (surrounded by leading and trailing ``--------''):

\textbf{-------- Problem begins}

\placeholder{problem\_statement}

\textbf{-------- Problem ends}

\mbox{}\\
Based on the above information, you will work in phases:

\mbox{}\\
\textbf{Phase 1: Define Stress Test Specifications.}\\
-- Based on the problem statement, describe the size, shape, and range of potential inputs.\\
-- Identify input patterns that can maximize the inefficiencies of the ``slow'' programs.\\
-- Don't just limit the generator into generating tests that maximize input size, but suggest specific values or patterns for inputs that can stress the slow programs (e.g., sorted arrays, repeated elements, specific edge cases).\\
-- Explain how the inputs should be designed to exploit algorithmic inefficiencies or implementation anti-patterns.

\mbox{}\\
\textbf{Phase 2: Implement Test Generator}\\
-- Write a Python script to generate the tests described in the previous phase.\\
-- The script should:\\
\hspace*{1em}-- Follow the input format specified in the problem statement. Note that \texttt{10000 (10\^5)} is usually written as \texttt{105}, \texttt{100000 (10\^6)} is usually written as \texttt{106}, etc.\\
\hspace*{1em}-- Maximize stress on the programs to make those inefficient programs get TLE or MLE.\\
\hspace*{1em}-- Avoid using random generator with a large range of values when generating the size input (e.g., length of array, etc.), e.g., \texttt{random.randint(1, 10000)} (say 10000 is the upper bound), as it may not stress the program enough. Instead, use values that are more likely to cause inefficiencies, for example, values that are same or closer to the upper bound, like \texttt{10000}, \texttt{9999}, \texttt{9990} or \texttt{random.randint(9900, 10000)}, etc. Feel free to directly use the upper bound as the size of the input.\\
\hspace*{1em}-- For numbers that are not the size of the input, generate random values to improve input diversity and cover different patterns.\\
\hspace*{1em}-- The test generator should generate a total of \placeholder{number\_of\_tests} test cases.\\
\hspace*{1em}-- Read an argument that specifies a directory and write all test cases as \texttt{'test\_01.in'}, \texttt{'test\_02.in'}, etc., into the directory. E.g., executing \texttt{python gen.py \placeholder{output\_directory}} will produce \texttt{'test\_01.in'}, \texttt{'test\_02.in'}, etc., in the \placeholder{output\_directory} directory.

}% end parbox
}% end fbox

\end{document}